\documentclass[12pt]{spieman}  
\usepackage{tocloft}
\usepackage{amsmath,amsfonts,amssymb}
\usepackage{mathabx}
\usepackage{deluxetable}
\usepackage{graphicx}
\usepackage{setspace}
\usepackage{float}
\usepackage{hhline}

\newcommand{\rom}[1]{\uppercase\expandafter{\romannumeral #1\relax}}

\def\gappeq{\mathrel{ \rlap{\raise.5ex\hbox{$>$}}
                      {\lower.5ex\hbox{$\sim$}}  } }

\title{The ExoEarth Yield Landscape for Future Direct Imaging Space Telescopes}

\author[a]{Christopher C. Stark}
\author[b]{Rus Belikov}
\author[c]{Matthew R. Bolcar}
\author[d]{Eric Cady}
\author[d]{Brendan P. Crill}
\author[e]{Steve Ertel}
\author[c]{Tyler Groff}
\author[d]{Sergi Hildebrandt}
\author[d]{John Krist}
\author[d]{P. Douglas Lisman}
\author[d]{Johan Mazoyer\footnote{NHFP Sagan Fellow}}
\author[d]{Bertrand Mennesson}
\author[f]{Bijan Nemati}
\author[a]{Laurent Pueyo}
\author[c]{Bernard J. Rauscher}
\author[d]{A.~J. Riggs}
\author[d]{Garreth Ruane}
\author[d]{Stuart B. Shaklan}
\author[b]{Dan Sirbu}
\author[a]{Remi Soummer}
\author[a]{Kathryn St. Laurent}
\author[c]{Neil Zimmerman}

\affil[a]{Space Telescope Science Institute, Baltimore, MD 21218, USA}
\affil[b]{NASA Ames Research Center, Moffett Field, CA 94035, USA}
\affil[c]{NASA Goddard Space Flight Center, Greenbelt, MD 20771, USA}
\affil[d]{Jet Propulsion Laboratory, Pasadena, CA 91109, USA}
\affil[e]{Large Binocular Telescope Observatory, Tucson, AZ 85721, USA}
\affil[f]{University of Alabama in Huntsville, Huntsville, AL 35899, USA}

\cftpagenumbersoff{figure}
\cftpagenumbersoff{table} 
\begin{document} 
\maketitle

\begin{abstract}

The expected yield of potentially Earth-like planets is a useful metric for designing future exoplanet-imaging missions.  Recent yield studies of direct-imaging missions have focused primarily on yield methods and trade studies using ``toy" models of missions.  Here we increase the fidelity of these calculations substantially, adopting more realistic exoplanet demographics as input, an improved target list, and a realistic distribution of exozodi levels.  Most importantly, we define standardized inputs for instrument simulations, use these standards to directly compare the performance of realistic instrument designs, include the sensitivity of coronagraph contrast to stellar diameter, and adopt engineering-based throughputs and detector parameters.  We apply these new high-fidelity yield models to study several critical design trades: monolithic vs segmented primary mirrors, on-axis vs off-axis secondary mirrors, and coronagraphs vs starshades.  We show that as long as the gap size between segments is sufficiently small ($<0.1\%$ of telescope diameter), there is no difference in yield for coronagraph-based missions with monolithic off-axis telescopes and segmented off-axis telescopes, assuming that the requisite engineering constraints imposed by the coronagraph can be met in both scenarios.  We show that there is currently a factor of $\sim$2 yield penalty for coronagraph-based missions with on-axis telescopes compared to off-axis telescopes, and note that there is room for improvement in coronagraph designs for on-axis telescopes. We also reproduce previous results in higher fidelity showing that the yields of coronagraph-based missions continue to increase with aperture size while the yields of starshade-based missions turnover at large apertures if refueling is not possible. Finally, we provide absolute yield numbers with uncertainties that include all major sources of astrophysical noise to guide future mission design.

\end{abstract}

\keywords{telescopes --- methods: numerical --- planetary systems}

\begin{spacing}{2}

\section{Introduction}
\label{intro}

In the study of almost any astronomical object, sample size is crucial.  Our ability to learn about objects hinges on comparison with other, similar objects.  Thus, the future scientific conclusions that we will be able to make with respect to habitability, life as we know it, and our place in the universe is critically linked to the sample size of potentially Earth-like planets studied by future direct-imaging space telescopes.

We can see the need for large sample sizes in exoplanet characterization studies already.  Over the last two decades we have realized the amazing diversity of extrasolar planets, which are often unlike anything in the Solar System\cite{rogers2015,fulton2017}.  This diversity appears to extend even to sub-populations of exoplanets; spectra of hot Jupiter atmospheres alone exhibit a wide range of features\cite{sing2016}. Understanding any one of these planets in detail requires a statistical analysis of a larger related population. Spectroscopic analysis of transiting hot Jupiter atmospheres have been primarily limited by sample size, motivating a large collaborative effort to increase the sample size to several dozen, one of the largest \emph{Hubble} Space Telescope projects ever conducted\cite{sing2016_2}.

Many previous studies have estimated the exoEarth candidate yields of future missions using the completeness methods first established by Ref.~\citenum{brown2005}.  Some of these studies were of relatively high fidelity in support of specific mission concepts \cite{savransky2010,savransky2016}.  Others focused on advancing the methods of yield calculation \cite{hunyadi2007,brown2010,stark2014_2,stark2016_2}.  Most recently these new methods were used to investigate broader trades, the sensitivity of yield to different mission parameters, and compare different exoplanet imaging instruments \cite{stark2015,stark2016}.  While the lessons learned from these analyses were valuable, they were based on relatively simple models of missions.

In this paper, we build on previous yield studies and advance the fidelity of the mission simulations.  We adopt more realistic astrophysical inputs, including a continuous distribution of exoplanet occurrence rates with estimated uncertainties, an updated target list, a more thorough treatment of stray light from binary stars, and a realistic exozodi distribution.  Most significantly, we dramatically advance the fidelity of the instrument simulations; we use detailed optical simulations of realistic coronagraph and starshade designs, include the contrast degradation due to stellar diameter, adopt end-to-end throughputs based on the LUVOIR Architecture A coronagraph design \cite{luvoir_interim}, and use a realistic detector model assuming reasonable future progress.  We then use these high fidelity inputs to estimate the exoEarth candidate yield of future missions, specifically addressing several major design trades: monolithic mirrors vs.~segmented mirrors, on-axis vs.~off-axis secondary mirrors, and coronagraphs vs.~starshades. We note that we do not address hybrid missions, in which coronagraphs and starshades are used together.

\section{Updates to astrophysical assumptions}

We maintained most of the astrophysical assumptions, including the zodiacal and exozodiacal brightness models, as in Ref.~\citenum{stark2015}.  Below we detail updates to the target list, treatment of binary stars, exoplanet distribution/occurrence rates, and exozodi distribution.

\subsection{Target list}

The potential target list of Ref.~\citenum{stark2015} was generated using the original Hipparcos catalog \cite{hip1997}.  Here, we adopted a new potential target list with more accurate stellar parameters and increased completeness.  

One of the most important parameters for accurate exoEarth completeness calculations is the distance to the star, as it sets the stellar luminosity as well as the absolute and angular scale of the habitable zone.  Ideally we'd have a vetted catalog of all nearby stars and GAIA DR2 parallax measurements for each one.  At the time of publication, such a catalog does not exist.  In its absence, we generated our own target list that simplistically approximates this.

We required that all stars have either \emph{Hipparcos}- \cite{hip1997,vanleeuwen2007} or GAIA-measured\cite{gaiadr1,marrese2018} parallaxes.  As such, we formed our potential target list from the union of the \emph{Hipparcos} New Reduction catalog\cite{vanleeuwen2007} and the GAIA TGAS catalog\cite{gaiadr1} (which correlates GAIA DR1 measurements with \emph{Hipparcos} and Tycho-2 designated stars), adopting the GAIA DR1 parallax measurements over \emph{Hipparcos} when available.  We used the apparent magnitudes and colors from the original \emph{Hipparcos} catalog\cite{hip1997}.  We then updated the distances of all stars that have been correlated with the GAIA DR2\cite{marrese2018} and down-selected to stars within 50 pc. Using SIMBAD we retrieved any missing UBVRIJHK photometry and correlated stars with the Washington Double Star catalog.  Finally, we used the Washington Double Star catalog to record companion magnitudes and separations, where available.  This procedure can be summarized as:

\begin{enumerate}
\item Start with full \emph{Hipparcos} New Reduction catalog \cite{vanleeuwen2007} and the photometry from the original \emph{Hipparcos} catalog \cite{hip1997}
\item Complement target list and update distances with the GAIA TGAS catalog\cite{gaiadr1}
\item Update distances of all stars correlated with GAIA DR2\cite{marrese2018}
\item Down-select to stars within 50 pc
\item Using SIMBAD, complement missing photometry and missing spectral types
\item Using SIMBAD, correlate targets with Washington Double Star catalog and record companion parameters when available 
\end{enumerate}

Like Ref.~\citenum{stark2015}, we then removed stars that have missing luminosities, missing V  magnitudes, and missing temperature classifications.  We required that stars have at least one magnitude in a band at wavelengths shorter than V, and at least one magnitude at wavelengths longer than V, such that we can reliably interpolate their spectra.  We also removed all giant stars (luminosity classes other than IV and V).  Unlike Ref.~\citenum{stark2015}, we retained all binary stars (see Section \ref{binary_section}).

The above process produces a target list that is nearly complete to $V\sim8$, and has improved distance measurements and more accurate photometry than the target list adopted by Ref.~\citenum{stark2015}.  Admittedly, this target list is still somewhat crude; any actual mission planning real observations would benefit greatly from a more thoroughly vetted target list.  However, the imperfections in our target list have a negligible impact on the yield results presented herein.  We note that the down-select to 50 pc does not impact our yield calculations of potentially Earth-like planets. Even the most capable simulated missions rarely select stars beyond 30 pc\cite{luvoir_interim}, due to a combination of HZ scale compared to the IWA, the planet brightness falling below the noise floor, and the long exposure times for planets above the noise floor around more distant stars.

To demonstrate the robustness of our results to target list imperfections, we compared yield calculations for a 4 m-diameter ``toy" coronagraph mission while intentionally varying the target list.  Keeping all other parameters constant, we compared the exoEarth candidate yield estimates for six target lists of varying precision.  One of these target lists was simply the original \emph{Hipparcos} catalog\cite{hip1997}, three were generated generated by taking the output of the above procedure at steps 1, 2, and 7, and the other two target lists were the highly detailed ExoCat-1 \cite{turnbull2015} and an intentionally unrefined catalog generated by a simple SIMBAD query.  We found that all yield results varied by less than 5\%.  Among the more refined target lists, variations were less than 2\%.

For smaller scale missions with very limited targets (less than a few dozen), the details of the potential target list could impact the expected yield to a greater degree.  However, all simulated missions in this paper do not fall in this category.

Unlike previous yield studies, we included the impact of stellar diameters on coronagraph contrast for the first time.  To do this, we estimated stellar diameters from $B-V$ color using Eq. 2 and Table 1 from Ref.~\citenum{boyajian2014}.  These equations are valid for luminosity class IV and V stars (all stars in our potential target list).  The equations have an empirically measured precision better than 8\%, such that uncertainties in stellar diameter have a very small impact on the raw contrast.  Because the uncertainty in stellar diameters is roughly Gaussian these small uncertainties in raw contrast should average out over the dozens of stars observed such that the impact on yield is negligible.  Ultimately, as we will show later, the best performing coronagraphs that we select are relatively insensitive to stellar diameter, such that the uncertainty on any single star's diameter has a negligible effect on its raw contrast.

\subsection{Binary stars and stray light\label{binary_section}}

Detecting an Earth-like planet around a Sun-like star requires suppressing starlight by a factor of 10 billion, creating a ``dark zone" within which exoplanets can be observed.  Thus, exposure times can be greatly impacted by what would normally be considered very low levels of background light.  One such source is stray light from nearby stars.  Even if a companion star is outside of the instrument's field of view, imperfections in or contamination of the mirrors can diffract light from the companion's PSF core into the far wings of its PSF, potentially flooding the instrument's dark zone with unwanted stray light.

The level of stray light depends on the separation and magnitude of the nearby star as well as the diffractive properties of the mirrors.  For each star in our potential target list, we searched for an entry in the Washington Double Star catalog via correlations resolved by SIMBAD.  We recorded the separation and magnitude of all companion stars.  We then calculated the flux from the stray light assuming a simple prescription for the wings of the PSF, as detailed in Section \ref{exposure_time_section}.

We included this stray light as a source of background and treated it similarly to all other astrophysical sources of background.  We assumed that it contributes to the photon noise only and can be modeled and subtracted from the final image.  I.e., we assumed that the stray light can be integrated over given the requisite exposure time.  The choice as to whether a binary system is worth observing given the level of stray light is ultimately made by the yield code during the observation optimization\cite{stark2014_2}. As expected, when including stray light many stars with companions are reduced in terms of observation priority or go completely unobserved.

\begin{figure}[H]
\centering
\includegraphics[width=3in]{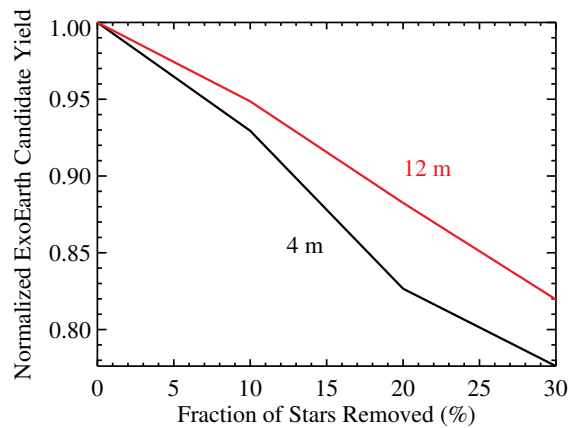}
\caption{Normalized exoEarth candidate yield as a function of the fraction of stars randomly removed for a toy 4 m mission (black) and a toy 12 m mission (red). Any refinements to the target list that remove X\% of stars will have a smaller than X\% impact on yield.\label{yield_vs_target_list_completeness_fig}}
\end{figure}

We note that any binarity not included in the Washington Double Star catalog is currently ignored.  There are undoubtedly stars in our target list with problematic companions that are not modeled.  To understand the impact of these missing binaries, we performed a rudimentary yield calculation for 4 m- and 12 m-diameter ``toy" missions: we calculated the exoEarth candidate yield while randomly removing larger and larger fractions of the target list.  As shown in Figure \ref{yield_vs_target_list_completeness_fig}, a 30\% reduction in the target list would reduce exoEarth candidate yield by $\sim$20\%.  The reduction in yield is not equal to the fraction of targets removed because as targets are removed, the exposure time that was devoted to those targets can be redistributed to other non-problematic targets.  Any future refinements to the target list that remove some fraction of the stars should follow a similar trend.

\subsection{Exozodi distribution\label{exozodi_section}}

Previous yield estimates adopted either a fixed exozodi level for every target or a distribution of exozodi that was poorly constrained by observations.  Recently, the Hunt for Observable Signatures of Terrestrial Systems (HOSTS) survey for exozodiacal dust using the Large Binocular Telescope Interferometer (LBTI) significantly improved our constraints on the exozodi distribution around Sun-like stars\cite{ertel2018,LBTI_report,ertel2019}.  The result is a best-fit free-form distribution to the observations with a median exozodi level of 4.5 zodis (nominal case), along with $\pm$1$\sigma$ distributions with median exozodi levels of 3 and 12 zodis (pessimistic and optimistic cases, respectively)\cite{LBTI_report,ertel2019}. We adopt the nominal distribution, and randomly assign exozodi levels from it to individual stars. The optimistic and pessimistic scenarios are included when estimating yield uncertainties, discussed in Section \ref{results_section}. The exozodi model describing surface brightness as a function of wavelength, circumstellar distance, and spectral type can be found in Reference \citenum{stark2014_2}.

\subsection{ExoEarth candidate definition and occurrence rates\label{eec_def}}

Most exoEarth yield estimates published thus far\cite{brown2005,hunyadi2007,brown2010,savransky2010,stark2014_2,stark2015,stark2016,stark2016_2} have adopted a simple, but inconsistent model for exoEarth candidates: an Earth twin with a given occurrence rate, $\eta_{\Earth}$, spread over a habitable zone (HZ) of a given size.  In reality, exoEarth candidates will cover a range of radii, and this range is directly tied to the value of $\eta_{\Earth}$.  Thus, we must more carefully define the boundaries of an exoEarth candidate and calculate $\eta_{\Earth}$ self-consistently. We adopt the same definition of an exoEarth candidate and the same occurrence rates as in the HabEx and LUVOIR interim reports \cite{habex_interim,luvoir_interim}.  Here we briefly summarize these assumptions.

The \emph{Kepler} mission revolutionized exoplanet demographics by detecting thousands of new planets with a wide range of radius and stellar insolation, including a handful of Earth-sized planets in the habitable zones of FGK stars.  However, the prime mission ended before \emph{Kepler} could discover a statistically significant sample of potentially Earth-like planets, requiring extrapolation to estimate $\eta_{\Earth}$ for Sun-like stars\cite{burke2015}.  As a result, published estimates of $\eta_{\Earth}$ for FGK stars vary significantly \cite{catanzarite2011,youdin2011,traub2012,petigura2013,foremanmackey2014,burke2015,mulders2018}.  To help achieve a useful community-wide consensus of occurrence rates for FGK stars, the NASA funded Exoplanet Exploration Program Analysis Group (ExoPAG) led Study Analysis Group 13 (SAG13).  SAG13 standardized a grid of period and planet radius, interpolated most published occurrence rates to this grid, collected additional contributions from the community, and compiled the results to form a community-wide average with uncertainties\cite{sag13_report,kopparapu2018}.  Because submissions came from similar data sets and the very nature of extrapolation does not adequately inform statistical significance, the uncertainties can be thought of as ``reasonably 1$\sigma$-like," as they include the majority of submitted values and extrapolation methods.  The SAG13 community-wide average occurrence rate grid for FGK stars is well-fit by a broken power law of the form
\begin{equation}
\frac{\partial^2 N(R_{\rm p},P)}{\partial\ln{R_{\rm p}}\; \partial\ln{P}} = \Gamma \left(\frac{R_{\rm p}}{{R}_{\Earth}}\right)^{\!\alpha} \left(\frac{P}{1\; {\rm yr}}\right)^{\!\beta},
\end{equation}
where $R_{\rm p}$ is planet radius and $P$ is orbital period.  The fit parameters for the expected value are given by
\begin{equation}
   [\Gamma, \alpha, \beta] = 
	\begin{cases}
	[0.38,-0.19,0.26], & \text{for}\; R_{\rm p} < 3.4 R_{\Earth} \\
	[0.73,-1.18,0.59], & \text{for}\; R_{\rm p} \ge 3.4 R_{\Earth}.
	\end{cases}
\end{equation}
For the upper bound on the occurrence rates,
\begin{equation}
   [\Gamma_+, \alpha_+, \beta_+] = 
	\begin{cases}
	[1.06,-0.68,0.32], & \text{for}\; R_{\rm p} < 3.4 R_{\Earth} \\
	[0.78,-0.82,0.67], & \text{for}\; R_{\rm p} \ge 3.4 R_{\Earth},
	\end{cases}
\end{equation}
and for the lower bound,
\begin{equation}
   [\Gamma_-, \alpha_-, \beta_-] = 
	\begin{cases}
	[0.138,0.277,0.204], & \text{for}\; R_{\rm p} < 3.4 R_{\Earth} \\
	[0.72,-1.58,0.51], & \text{for}\; R_{\rm p} \ge 3.4 R_{\Earth}.
	\end{cases}
\end{equation}

To determine a value of $\eta_{\Earth}$, we must integrate the above fit over some radius-period bin that defines an exoEarth candidate. We defined the period boundaries in terms of semi-major axis for a Sun-like star and adopted the conservative HZ defined by Ref.~\citenum{kopparapu2013} and \citenum{kopparapu2014}.  Climate models performed in 3D suggest a slight inward revision to the warm edge of the HZ\cite{leconte2013}, such that our final adopted semi-major axis range is $0.95$--$1.67$ AU for a solar twin and scales with the square root of the stellar bolometric luminosity $L_{\star}$.

For planet radius boundaries, we adopted an upper limit of $1.4$ ${\rm R}_{\Earth}$, motivated by the transition from rocky to gas-dominated planets\cite{rogers2015}.  Note that the exact value of this upper boundary on radius does not significantly impact the yield, as the occurrence rates we adopted are weighted toward smaller planets (see below).  For the lower boundary on planet radius, we adopted a value of $0.8((a/1\; {\rm AU}))^{-0.5}$ $R_{\Earth}$, motivated by an empirical relationship in the solar system that describes which bodies are able to retain significant atmospheres\cite{zahnle2017}.  While this lower limit has a significant impact on $\eta_{\Earth}$, ultimately it has a smaller impact on exoEarth candidate yield, as the smallest planets take too long to detect/characterize or fall below the assumed noise floor.  

\begin{figure}[H]
\centering
\includegraphics[width=3in]{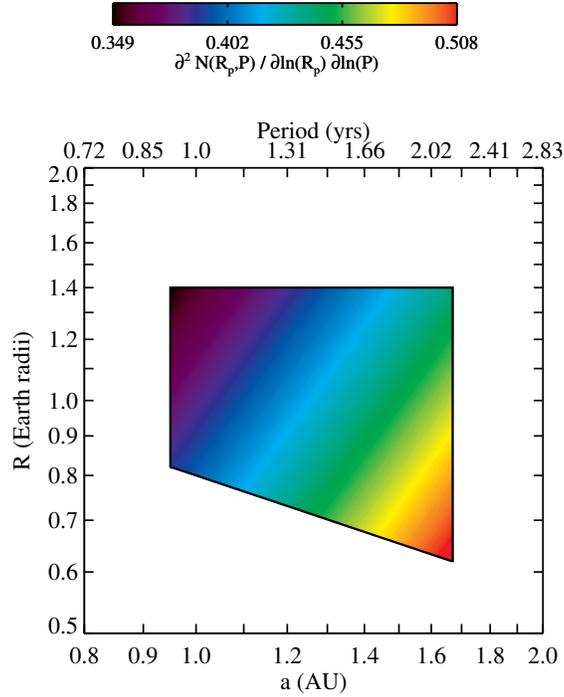}
\caption{The SAG13 occurrence rates bounded by our exoEarth candidate definition. Integrated over the bin, $\eta_{\Earth}= 0.24^{+0.46}_{-0.16}$, but not all of these will be detectable since the planet distribution is weighted toward fainter planets. Semi-major axis boundaries are set by the conservative HZ\cite{kopparapu2013,leconte2013,kopparapu2014}, the upper radius boundary is set by the transition from rocky to gas-dominated planets\cite{rogers2015}, and the lower radius boundary is set by atmospheric escape\cite{zahnle2017}. \label{planet_distribution_fig}}
\end{figure}

Integrating the SAG13 occurrence rates over these boundaries gives $\eta_{\rm \Earth} = 0.24^{+0.46}_{-0.16}$.  This expected $\eta_{\Earth}$ value is a factor of $2.4$ times larger than was assumed in Ref.~\citenum{stark2015}.  However, this has less than a factor of 2 impact on yield because, again, smaller planets dominate the distribution.  Figure \ref{planet_distribution_fig} shows the exoEarth candidate boundaries and the SAG13 distribution of planets within.

Because the SAG13 results are averaged over FGK stars, there is no spectral type dependence expressed in the fit.  As a result, we must either assume the occurrence rates scale with the HZ such that all stars have the same value of $\eta_{\Earth}$, or that the occurrence rates are to be evaluated for each star independently based on the planet's radius range and the HZ's absolute semi-major axis range.  The latter would implicitly introduce a spectral type dependence, such that the effective $\eta_{\Earth}$ for F stars would be a factor of $1.5$ times greater than that of K stars.  Given that no Earth-sized planets have been detected in the HZ of F or G type stars, there is no direct evidence for this spectral type dependence.  Thus, we chose to avoid biasing the yield calculations toward earlier type stars, and scaled the occurrence rates with the HZ, such that all stars have the same $\eta_{\Earth}$.  We note that we also applied the SAG13 occurrence rates to M stars, a valid approximation, as the occurrence rates of Ref.~\citenum{dressing2015} integrated over our definition of an exoEarth candidate provides a similar value of $\eta_{\Earth}$.

Another primary source of uncertainty in our definition of an exoEarth candidate is the albedo distribution of the planets.  We expect this quantity to be immeasurable prior to the launch of a direct imaging mission like those simulated here.  Therefore we maintain the assumption of Ref.~\citenum{stark2014_2} and adopt Earth's geometric albedo of $0.2$ at all wavelengths simulated (see, e.g., Ref.~\citenum{roberge2017}).   We also maintain the assumption of a Lambertian phase function.  Table \ref{exoearth_params_table} summarizes our astrophysical assumptions and the parameters that define an exoEarth candidate.

\begin{deluxetable}{ccl}
\tablewidth{0pt}
\footnotesize
\tablecaption{Baseline Astrophysical Parameters\label{exoearth_params_table}}
\tablehead{
\colhead{Parameter} & \colhead{Value} & \colhead{Description} \\
}
\startdata
$\eta_{\Earth}$ & $0.24$ & Fraction of Sun-like stars with an exoEarth candidate\tablenotemark{a} \\
$R_{\rm p}$ & $[0.6,1.4]$ $R_{\Earth}$ & ExoEarth candidate radius range\tablenotemark{a} \\
$a$ & $[0.95,1.67]$ AU & ExoEarth candidate semi-major axis range\tablenotemark{b} \\
$e$ & $0$ & Eccentricity (circular orbits) \\
$\cos{i}$ & $[-1,1]$ & Cosine of inclination (uniform distribution) \\
$\Omega$ & $[0,2\pi)$ & Argument of pericenter (uniform distribution) \\
$M$ & $[0,2\pi)$ & Mean anomaly (uniform distribution) \\
$\Phi$ & Lambertian & Phase function \\ 
$A_G$ & $0.2$ & Geometric albedo of exoEarth candidate at $0.55$ and $1$ $\mu$m \\
$z$ & 23 mag arcsec$^{-2}$\tablenotemark & Average V band surface brightness of zodiacal light\tablenotemark{c} \\
$z'$ & 22 mag arcsec$^{-2}$\tablenotemark  & V band surface brightness of 1 zodi of exozodiacal dust\tablenotemark{d} \\
$n$ & $4.5$ & Median exozodi level\tablenotemark{e} \\
\enddata
\vspace{-0.1in}
\tablenotetext{a}{See Section \ref{eec_def}}
\tablenotetext{b}{For a solar twin.  The habitable zone is scaled by $\sqrt{L_{\star}/L_{\Sun}}$.}
\tablenotetext{c}{Varies with ecliptic latitude.}
\tablenotetext{d}{For Solar twin. Varies with spectral type and planet-star separation---see Appendix C in Ref.~\citenum{stark2014_2}.}
\tablenotetext{e}{Individual exozodi levels randomly drawn from LBTI HOSTS best-fit distribution}
\end{deluxetable}

\section{Exposure Time Calculation\label{exposure_time_section} and Mission Parameters}

While we adopt the same basic framework for exposure time calculation as in Ref.~\citenum{stark2014_2}, we slightly modify how some of the factors are determined.  For clarity, here we reproduce many of the exposure time equations from Ref.~\citenum{stark2014_2} with updated notation.  

The exposure time of any individual planet was calculated according to
\begin{equation}
\label{tau_equation}
	\tau = \left({\rm S/N}\right)^2 \left(\frac{{\rm CR_p} + 2\, {\rm CR_b} }{{\rm CR_p}^2}\right),
\end{equation}
where $\rm{S/N}$ is the required signal to noise ratio, ${\rm CR_p}$ is the count rate for the planet, ${\rm CR_b}$ is the count rate for all sources of background, and the factor of 2 assumes some sort of background subtraction.  The count rate of the planet is given by a slightly modified equation,
\begin{equation}
\label{CRp_equation}
	{\rm CR_p} = F_0\, 10^{-0.4\left({\rm m}_{\lambda} + \Delta {\rm mag_p}\right)}\, A\, \Upsilon_{\rm c}\!\left(x,y\right)\, T\, \Delta \lambda,
\end{equation}
where $F_0$ is the zero-magnitude flux at the wavelength of interest $\lambda$, $m_{\lambda}$ is the stellar apparent magnitude at $\lambda$, $\Delta{\rm mag_p}$ is the magnitude difference between the planet and star, $\Delta\lambda$ is the bandpass, and $A$ is the effective collecting area of the telescope aperture accounting for segment gaps and secondary mirror and strut obscurations. The notable changes here are simply in the definitions of throughput.  Previously $\Upsilon$ was simply the fraction of the planet's PSF in the photometric core, and was set to a uniform value of $0.69$.  We replaced this with $\Upsilon_{\rm c}\!\left(x,y\right)$, the coronagraph's spatially-dependent core throughput, which is defined as the fraction of light entering the coronagraph that ends up in the photometric core of the planet's PSF assuming perfectly reflecting/transmitting optics.  Because $\Upsilon_{\rm c}\!\left(x,y\right)$ now includes the intentional diffractive and absorptive properties of the coronagraph masks, it is typically significantly less than $0.69$.  Moreover, $\Upsilon_{\rm c}\!\left(x,y\right)$ is no longer a simple assumption---it is the product of detailed coronagraph simulations, to be discussed in Section \ref{coronagraph_section}.  Accordingly, $T$ is now the non-coronagraphic end-to-end throughput, including everything except the coronagraph's core throughput, discussed in further detail in Section \ref{throughput_section}.

The leaked stellar count rate expression was also slightly modified to
\begin{equation}
\label{CRbstar_equation}
	{\rm CR_{b,\star}} = F_0\, 10^{-0.4{\rm m}_{\lambda}}\, \frac{I\!\left(x,y\right)}{\theta^2} \Omega\, A\, T\, \Delta \lambda,
\end{equation}
where we have replaced the factors $\zeta\,{\rm PSF_{\rm peak}}$ in Ref.~\citenum{stark2014_2} with $I\!\left(x,y\right)/\theta^2$.  Here $I\!\left(x,y\right)/\theta^2$ is the spatially dependent leaked stellar count rate per unit solid angle exiting the instrument normalized to the starlight entering the instrument, and $\Omega$ is the solid angle of the photometric aperture used for planet detection.  The reason for this change of notation is that in practice, $I\!\left(x,y\right)$ is a simulated 2D image with pixel size $\theta$, calculated via detailed instrument simulations.  

The zodiacal and exozodiacal background count rates are given by
\begin{equation}
\label{CRbzodi_equation}
	{\rm CR_{b,zodi}} = F_0\, 10^{-0.4z}\, \Omega\, A\, T\, T_{\rm sky}\!\left(x,y\right)\, \Delta \lambda,
\end{equation}
and
\begin{equation}
\label{CRbexozodi_equation}
	{\rm CR_{b,exozodi}} = F_0\, n\, 10^{-0.4z'\!\left(x,y\right)}\, \Omega\, A\, T\, T_{\rm sky}\!\left(x,y\right)\, \Delta \lambda,
\end{equation}
where $z$ is the surface brightness of the zodiacal light in magnitudes per unit solid angle, calculated at the desired wavelength and nominal pointing to the desired target to include the spatial variation of the zodiacal cloud's surface brightness (see Appendix B in Ref.~\citenum{stark2014_2} and Ref.~\citenum{stark2016}), $z'$ is the surface brightness of 1 zodi of exozodiacal light in magnitudes per unit solid angle, and $n$ is the number of zodis assumed for all stars (details on our treatment of exozodiacal surface brightness can be found in the appendices of Ref.~\citenum{stark2014_2}).  Here the definition of throughput was again changed.  $T_{\rm sky}\!\left(x,y\right)$ approximates the instrument's throughput for extended sources.  Ideally we'd convolve our 2D exozodi models with the instrument's spatially-dependent PSF for each star individually.  However, this would be numerically taxing and our disk model is an approximation that assumes no specific orientation.  Therefore, we approximate the instrument's effects on our disk model by first convolving the spatially-dependent PSF at all locations with a normalized uniform background, then simply multiply each disk model by the resulting throughput factor, $T_{\rm sky}\!\left(x,y\right)$. 

We updated our calculation of the detector noise count rate from Equation 2 in Ref.~\citenum{stark2015} to include clock induced charge.  Clock induced charge is a noise term that becomes apparent when operating an EMCCD in Geiger mode (also called photon-counting mode). To minimize clock induced charge, one would use the longest possible photon counting time.  However, this comes at the cost of losing dynamic range for the brightest source in the scene, as two photons may arrive during a single frame.  There is therefore a tradeoff between the brightest astronomical object we wish to detect and the noise introduced.  

The Geiger efficiency, $q$, quantifies the probability that one or fewer photons arrive during a frame.  Letting CR$_{\rm sat}$ be the count rate of the brightest pixel for which we wish to achieve a given Geiger efficiency, we can express $q$ as
\begin{equation}
	q = \sum_{\gamma=0}^{1} P\!\left({\rm CR_{sat}}\, t\right),
\end{equation}
where $t$ is the time between frames and $P$ is the probability density function of the Poisson distribution. We can rewrite the above as
\begin{equation}
	q = \left(1 + {\rm CR_{sat}}\, t\right) e^{-{\rm CR_{sat}}\, t},
	\end{equation}
which has the solution
\begin{equation}
	t = -\frac{1}{\rm CR_{sat}} \left[1 + {\rm W}_{\!-1}\!\left(-\frac{q}{e}\right) \right],
\end{equation}
where ${\rm W}_{\!-1}$ is the lower branch of the Lambert W function and $e$ is Euler's number.  An EMCCD's clock induced charge, CIC, is expressed in units of counts pix$^{-1}$ frame$^{-1}$.  Thus, we can convert CIC to an effective dark current by dividing CIC by the above frame time:
\begin{equation}
	\xi' = \xi - {\rm CIC}\, {\rm CR}_{\rm sat} \left[1 + {\rm W}_{\!-1}\!\left(-\frac{q}{e}\right) \right]^{-1},
\end{equation}
where $\xi$ is the traditional dark current in units of counts pix$^{-1}$ s$^{-1}$.  Adopting $q=0.99$ such that we lose only 1\% of photons from the brightest pixel, the detector count rate becomes
\begin{equation}
 	{\rm CR}_{\rm b,detector} = n_{\rm pix} \left(\xi + {\rm RN}^2 / \tau_{\rm read} + 6.73\, {\rm CR}_{\rm sat}\, {\rm CIC}\right),
\end{equation}
where $n_{\rm pix}$ is the number of imager or integral field spectrograph (IFS) pixels per spectral element covered by the core of the planet's PSF, RN is the read noise, and $\tau_{\rm read}$ is the length of an individual read.  We set CR$_{\rm sat}$ equal to 10 times the count rate expected for a PSF core pixel of an Earth twin at quadrature, evaluated around each star individually.  For imaging, we set $n_{\rm pix} = 4$.  For $R=70$ spectral characterizations with the IFS at $\lambda_{\rm c} = 1$ $\mu$m, we set $n_{\rm pix} = 192$---we assumed each IFS lenslet spreads the light into 6 pixels per spectral element (3 in the spatial dimension and 2 in the spectral dimension), the PSF core covers 16 lenslets at 1 $\mu$m assuming Nyquist sampling at $0.5$ $\mu$m, and a factor of 2 increase in frequency sampling assuming a native IFS resolution of $R=140$.

To include the stray light from nearby stars discussed in Section \ref{binary_section}, we calculated the count rate via
\begin{equation}
	{\rm CR_{b,stray}} = F_0\, 10^{-0.4\left({\rm m}_{\lambda} + \Delta {\rm mag}_{\rm b}\right)}\, {\rm PSF}'\!\left(s_{\rm b}\right)\, \Omega\, A\, T\, \frac{T_{\rm sky}\!\left(x,y\right)}{\theta^2}\, \Delta \lambda,
\end{equation}
where $\Delta {\rm mag}_{\rm b}$ is the difference in magnitude between the star and binary companion, ${\rm PSF}'\!\left(s_{\rm b}\right)$ is the ratio of the PSF evaluated at a distance of $s_{\rm b}$ from the central peak, and $T_{\rm sky}\!\left(x,y\right)$ is assumed to have the same pixel scale $\theta$ as $I\!\left(x,y\right)$.  We calculated ${\rm PSF}'\!\left(s_{\rm b}\right)$ numerically assuming the PSF wings scale as $f^{-3}$, where $f$ is the spatial frequency of optical aberrations.

Unlike Refs.~\citenum{stark2014_2} and \citenum{stark2015}, we imposed a limit on exposure times.  We required all detection and spectral characterization times, including overheads, to be $<2$ months.  Any planets that did not meet this criteria did not count toward the yield.  While 2 months is quite long, in practice most planets fall well below this limit.  In reality, some planets may move behind the inner working angle or into a faint crescent phase during this time---such a planet would have to be spectrally characterized over multiple epochs. Given our adopted observation plan for coronagraph-based missions, in which we assume orbit determination prior to spectral characterization, and the fact that the overheads are far shorter than the spectral characterization times, breaking spectral characterizations into multiple visits should not impact our results significantly.  This assumption may be less valid for starshade-based missions.

\begin{deluxetable}{ccl}
\tablewidth{0pt}
\footnotesize
\tablecaption{Coronagraph-based Mission Parameters\label{baseline_params_table}}
\tablehead{
\colhead{Parameter} & \colhead{Value} & \colhead{Description} \\
}
\startdata
& & \bf{General Parameters} \\
$\Sigma \tau$ & $2$ yrs & Total exoplanet science time of the mission \\
$\tau_{\rm slew}$ & 1 hr & Static overhead for slew and settling time \\
$\tau_{\rm WFC}$ & $5 \left( \frac{A_0 \Upsilon_0 }{A \Upsilon}\right)$ hrs & Static overhead to dig dark hole (see Eq \ref{WFSC_equation})\\
$\tau'_{\rm WFC}$ & 1.1 & Multiplicative overhead to touch up dark hole \\
$X$ & $0.7$ & Photometric aperture radius in $\lambda/D_{\rm LS}$\tablenotemark{*} \\
$\Omega$ & $\pi(X\lambda/D_{\rm LS})^2$ radians & Solid angle subtended by photometric aperture\tablenotemark{*} \\
$\zeta_{\rm floor}$ & $10^{-10}$ & Raw contrast floor \\
$\Delta$mag$_{\rm floor}$ & $26.5$ & Noise floor (faintest detectable point source at S/N$_{\rm d}$) \\
$T_{\rm contam}$ & $0.95$ & Effective throughput due to contamination \\

\hline
& & \bf{Detection Parameters} \\
$\lambda_{\rm d,1}$ & $0.45$ $\mu$m & Central wavelength for detection in SW coronagraph \\
$\lambda_{\rm d,2}$ & $0.55$ $\mu$m & Central wavelength for detection in LW coronagraph \\
S/N$_{\rm d}$ & $7$ & S/N required for detection (summed over both coronagraphs) \\
$T_{\rm optical,1}$ & $0.16$/$0.57$\tablenotemark{**} & End-to-end reflectivity/transmissivity at $\lambda_{\rm d,1}$ \\
$T_{\rm optical,2}$ & $0.35$/$0.56$\tablenotemark{**} & End-to-end reflectivity/transmissivity at $\lambda_{\rm d,2}$ \\
$\tau_{\rm d,limit}$ & 2 mos & Detection time limit including overheads \\

\hline
& & \bf{Characterization Parameters} \\
$\lambda_{\rm c}$ & $1.0$ $\mu$m & Wavelength for characterization in LW coronagraph IFS \\
S/N$_{\rm c}$ & $5$ & Signal to noise per spectral bin evaluated in continuum \\
R & 70 & Spectral resolving power \\
$T_{\rm optical,IFS}$ & $0.21$/$0.30$\tablenotemark{**} & End-to-end reflectivity/transmissivity at $\lambda_{\rm c}$ \\
$\tau_{\rm c,limit}$ & 2 mos & Characterization time limit including overheads \\

\hline
& & \bf{Detector Parameters} \\
$n_{\rm pix,d}$ & 4 & \# of pixels in photometric aperture of imager at $\lambda_{\rm d,\#}$\\
$n_{\rm pix,c}$ & 192 & \# of pixels per spectral bin in LW coronagraph IFS at $\lambda_{\rm c}$ \\
$\xi$ & $3\times10^{-5}$ $e^-$ pix$^{-1}$ s$^{-1}$ & Dark current\\
RN & 0 $e^-$ pix$^{-1}$ read$^{-1}$ & Read noise\\
$\tau_{\rm read}$& N/A & Time between reads\\
CIC & $1.3\times10^{-3}$ $e^-$ pix$^{-1}$ frame$^{-1}$ & Clock induced charge\\
$T_{\rm QE}$ & $0.9$ & Raw QE of the detector at all wavelengths \\
$T_{\rm read}$ & $0.75$ & Effective throughput due to bad pixel/cosmic ray mitigation \\

\hline
\enddata
\vspace{-0.1in}
\tablenotetext{*}{$D_{\rm LS}$ is the diameter of Lyot stop projected onto the primary mirror}
\tablenotetext{**}{Smaller value is for TMA design with UV and VIS coronagraph channels, larger value is for Cassegrain design with two visible coronagraph channels.}

\end{deluxetable}

\subsection{Non-coronagraphic throughput\label{throughput_section}}

To estimate the exposure times required for exoplanet detection, we must have an accurate estimate of the end-to-end throughput of the system.  We define the non-coronagraphic end-to-end throughput as
\begin{equation}
	T\!\left(\lambda\right) = T_{\rm contam}\; T_{\rm optical}\!\left(\lambda\right) T_{\rm QE}\; T_{\rm read},
\end{equation}
where $T_{\rm contam}$ accounts for light lost due to contamination, $T_{\rm optical}$ is the optical throughput (the reflectivity/transmissivity of all optics), $T_{\rm QE}$ is the raw quantum efficiency (QE) of the detector, and $T_{\rm read}$ is an effective throughput factor resulting from bad pixel and cosmic ray mitigation.  For the contamination budget, we simply adopt $T_{\rm contam} = 0.95$ for all wavelengths, on par with the particulate coverage fraction for JWST's mirrors\cite{wooldridge2014}.  Below we describe our assumptions for the remaining throughput terms.

\subsubsection{Optical throughput}

For $T_{\rm optical}\!\left(\lambda\right)$, we calculated the reflectivity/transmissivity of every optic as a function of wavelength for realistic optical layouts.  All reflective optics were assumed to have coatings of protected aluminum, silver, or gold, depending on the bandpass of the channel in which they are used.  Reflectivities of these materials were obtained from measurements of representative samples\cite{palik1985,palik1991,quijada2012,quijada2014,sheikh2016}. Transmissive optics were assumed to be coated with high-performance broadband antireflection coatings, with an average transmissivity greater than $0.99$ for all wavelengths.  The transmissivities of the substrate materials were obtained from the material manufacturer.

\begin{figure}[H]
\centering
\includegraphics[width=6.5in,trim={0 2.6in 0 0},clip]{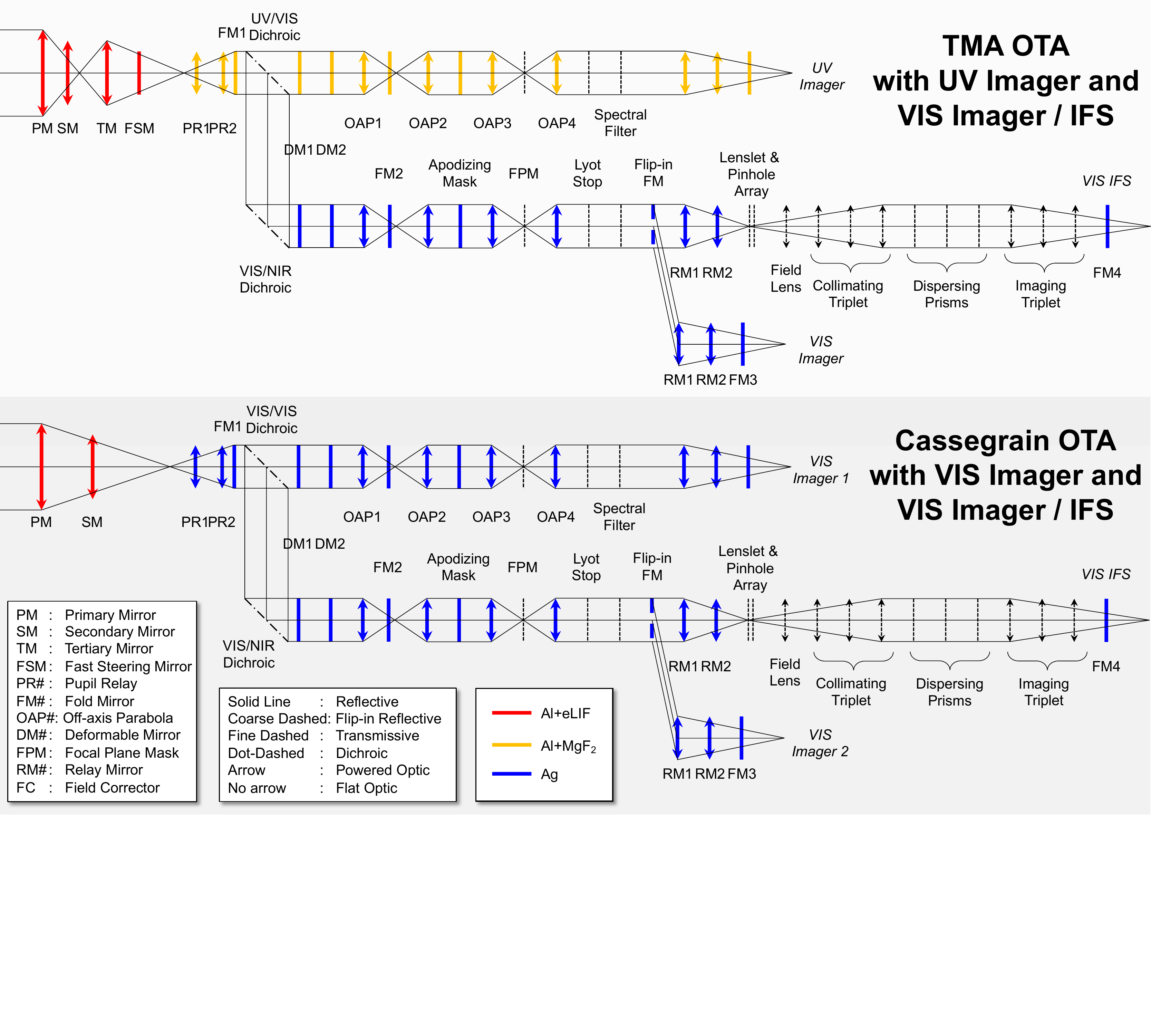}
\caption{Illustration of the two optical layouts considered with aluminum coated mirrors shown in red and orange.  The low and high throughput scenarios are shown on the top and bottom, respectively.  In the low throughput scenario, the VIS imager sees 7 aluminum coated mirrors. The high throughput scenario has only two aluminum coated mirrors, at the expense of UV coronagraphy.   \label{optical_layout_fig}}
\end{figure}

While it is incorrect to assume that all future direct imaging missions have the same optical layout and thus throughput, we can bound the problem by assuming two reasonable limiting scenarios. We adopted limits that stem from detailed coronagraphic optical layouts examined as part of the LUVOIR Architecture A study\cite{luvoir_interim}.  Both of these scenarios assume two coronagraph imagers operating in parallel, with an optional IFS in one channel for spectral characterization of exoplanets.  While neither of them exactly represents the design adopted for LUVOIR, they do represent plausible future designs.  These layouts are illustrated in Figure \ref{optical_layout_fig}.

The top panel in Figure \ref{optical_layout_fig} shows a low throughput scenario in which we assumed a maximum number of aluminum coated mirrors.  While aluminum coated mirrors are necessary to enable UV science, they have a relatively low reflectivity in the visible.  We assumed a three mirror anastigmat (TMA) optical telescope assembly (OTA) with a fourth fast steering mirror; all four mirrors are aluminum coated.  After the telescope, there are three additional pre-coronagraph optics.  In this scenario we assumed that one of the two coronagraph channels operates in the UV down to $\sim$300 nm (to enable the detection of ozone on potentially Earth-like planets), such that the three pre-coronagraph optics must also be aluminum coated.  The visible wavelength (VIS) coronagraph imager/IFS therefore sees seven aluminum coated mirrors.

The bottom panel in Figure \ref{optical_layout_fig} illustrates the high throughput scenario in which we assumed a minimum number of aluminum coated mirrors.  For this scenario, we adopted a Cassegrain telescope design with only two aluminum coated telescope mirrors, preserving UV science for other instruments in the observatory.  However, to examine a true upper limit on throughput, we removed the UV coronagraph capability and assumed that both coronagraph channels operate in the visible.  As a result, all three pre-coronagraph optics are silver coated.  

\begin{figure}[H]
\centering
\includegraphics[width=4in]{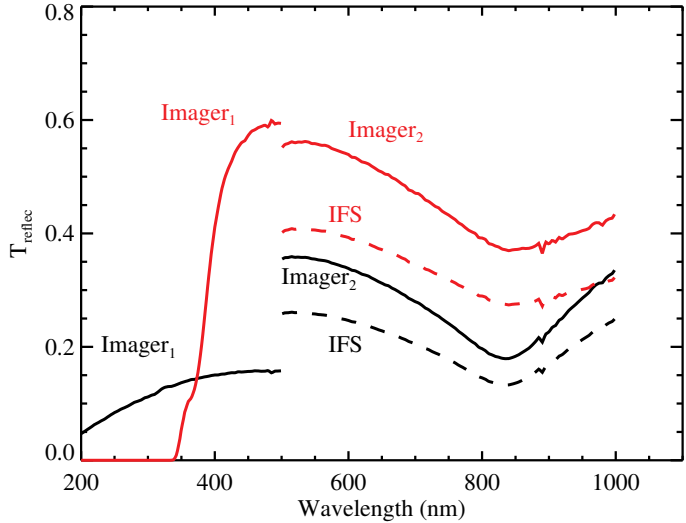}
\caption{End-to-end optical reflectivity/transmissivity for the low throughput (black) and high throughput (red) scenarios.  Each scenario assumes two coronagraph imagers (solid lines) operating in parallel, with a transition wavelength of 500 nm, and a visible wavelength IFS (dashed lines).  A UV coronagraph reduces the throughput of the entire coronagraphic system.  \label{throughput_curve_fig}}
\end{figure}

The black and red curves in Figure \ref{throughput_curve_fig} show $T_{\rm optical}\!(\lambda)$ for the low and high throughput scenarios, respectively.  For each scenario, we show the throughputs for the two coronagraph imagers (solid lines), along with the throughput of the coronagraph IFS (dashed lines).  Table \ref{baseline_params_table} summarizes $T_{\rm optical}$ integrated over the bandpasses used in this study.  We note that while we will use these two scenarios to illustrate how the optical telescope assembly (OTA) design and UV capability impact exoEarth candidate yield, ultimately the optical layout of the system is also driven by many factors not discussed here, including packaging, complexity, and the requirements of other instruments.

We note that both throughput scenarios assume a system designed to operate simultaneously on both polarization components.  Polarization aberrations can be minimized through careful design of the optical system, reducing the contrast leakage due to the differential and cross-polarization terms. Preliminary physical optics modeling of the coronagraph performance indicates that keeping the angle of incidence at any surface $\lesssim$12$^{\circ}$, and using fold mirrors in crossed-pairs only, does not prevent the coronagraph from achieving $10^{-10}$ contrast\cite{luvoir_interim}. If future studies determine that the DMs cannot simultaneously correct aberrations for orthogonal polarizations, a polarizer may be required, reducing the optical throughput by a factor of 2.  While we excluded this factor, we expect that it would result in a $\sim$20\% reduction in the exoEarth candidate yields presented herein based on typical yield scaling relationships.

\subsubsection{Detector efficiencies}

A primary science goal of these future missions is the search for water vapor in the atmospheres of potentially habitable planets. The most promising water vapor absorption feature in the Earth's spectrum occurs in the near-IR at 950 nm. Thus, future missions will require detectors with high near-IR QE, $\sim$90\% at 950 nm, and with very low noise. For comparison, the WFIRST EMCCD achieves a raw QE $\sim$10\% at 950 nm. This is a consequence of the silicon's thickness and the specific AR coating that is used (Teledyne-e2v has other CCD technologies and AR coatings that are better optimized for the near-IR). Nevertheless, future missions must advance detector technologies significantly.

Several recently developed or emerging $\lesssim$1 e- read noise detector technologies may prove viable in the near-IR. In silicon, these include p-channel CCDs with buried contacts (See Figs 2b and 3 of Ref.~\citenum{bebek2015}), the Hole Multiplying CCD \cite{rauscher2016}, and potentially thick, fully depleted skipper CCDs \cite{tiffenberg2017}. Superconducting technologies including microwave kinetic inductance devices (MKID) and transition edge sensor microcalorimeters (TES) may also be relevant, although these would require active cooling that brings its own challenges. Here we assume that appropriate detectors will eventually become available, discuss our adopted detector parameters, and compare with the performance of the WFIRST EMCCD.

We assumed the same detector properties for the UV imager, VIS imager, and VIS IFS.  For our baseline detector, we assumed 90\% raw QE at all wavelengths and adopted the noise parameters listed in Table \ref{baseline_params_table}.  For comparison, the WFIRST EMCCD beginning of life dark current is a factor of $\sim$5 worse than our assumptions, but has similar read noise and clock induced charge\cite{harding2016}.

In addition to the raw QE, there are effective QE terms associated with photon counting efficiency, mitigation of hot pixels and cosmic rays, and the charge transfer efficiency due to charge traps.  For the WFIRST EMCCD, these inefficiencies currently equate to another effective QE factor of $T_{\rm read} \approx 0.5$\cite{harding2016}.  Assuming plausible future improvement, primarily with respect to the charge transfer efficiency, we adopted $T_{\rm read} = 0.75$.


\subsection{Overheads}

Refs.~\citenum{stark2014_2} and \citenum{stark2015} assumed one year of total exposure time and one additional year for overheads, for a total of two years of total exoplanet science time.  Here we also assumed two total years of exoplanet science time, but we adopted a more informed treatment for overheads.  

We assume overheads can be separated into three dominant categories: slew, settle, and wavefront control times.  The slew overhead represents the time it takes to repoint the observatory to a new target, settling overheads represent the time it takes for the telescope to reach the dynamical and thermal equilibrium required to begin science exposures, and the wavefront control overheads represent the time it takes to dig the coronagraphic dark hole.  In reality, these overheads are not separable---the slew rate is chosen to reduce dynamic excitations of the observatory and minimize settling time, and the wavefront control can begin somewhat prior to the observatory reaching its final settled state.  Further, precise estimates of the overheads require detailed modeling of dynamical disturbances that are beyond the scope of this paper.  In light of this, we adopt conservative estimates based on known scaling relationships and estimates from current missions.

We adopted the same operational scenario baselined for the Wide Field Infrared Space Telescope Coronagraphic Instrument (WFIRST CGI). In this scenario, the observatory initially slews to a bright star nearby the target star, digs the dark hole on the bright star, repoints to the science star using the fine steering mirror (FSM), and touches up the dark hole with an iteration or two of the wavefront control algorithm.

First, we treated the slew and settle times as a single static overhead independent of aperture size.  For the James Webb Space Telescope (JWST), the slew and settle time is $0.5$ hours for a $53^{\circ}$ slew\cite{jwst_slew_jdox}.  While we expect the typical slew for our program to be significantly less, $\sim10^{\circ}$, we expect the settling time to take longer as the coronagraph places much stricter requirements on stability.  We thus adopt 1 hour in total for the typical slew and settle time, which we apply to every observation.  These assumed slew and settle times are reasonable, as has been demonstrated by the HabEx and LUVOIR mission concept studies, which adopt architectures specifically designed to minimize dynamic and thermal disturbances\cite{habex_interim, luvoir_interim}.

For the wavefront control overhead, we adopted values roughly consistent with the WFIRST CGI.  Although the exact time required to dig the dark hole using the WFIRST CGI is unknown, times $\sim10$ hours have been discussed to achieve $\sim10^{-9}$ contrast.  Because the wavefront control time scales with the photon detection rate, and because we assume future coronagraphic instruments with higher throughput and QE in dual polarization, this translates to roughly $5$ hours for a future off-axis monolithic 4 m telescope using a charge 6 vortex coronagraph at $10^{-10}$ contrast.  Thus, we adopt wavefront control overheads given by
\begin{equation}
\label{WFSC_equation}
	\tau_{\rm WFSC} = 5 \left( \frac{A_{0} \Upsilon_{\rm c,0} T_0 }{A \Upsilon_{\rm c} T}\right)\; {\rm hours} + 0.1\, \tau,
\end{equation}
where $A_0$ is the collecting area of an unobscured off-axis 4 m telescope, $\Upsilon_{\rm c,0}$ is the  core throughput of a charge 6 vortex coronagraph for an unobscured aperture, $T_0$ is the non-coronagraphic end-to-end throughput for our low throughput scenario, and $A$, $\Upsilon_{\rm c}$, and $T$ are the collecting area, coronagraphic core throughput, and non-coronagraphic end-to-end throughput of the telescope in study.  Given the shapes of the core throughput curves for different coronagraphs, to be discussed later, we evaluated the core throughputs at the inner working angle for vortex coronagraphs, and just beyond the inner working angle for apodized pupil Lyot coronagagraphs.

After the dark hole is generated to $10^{-10}$, the observatory repoints to the science target using the FSM.  We assumed this time is negligibly small.  Finally, we included the overheads for a single iteration of wavefront control, calculated as 10\% of the science exposure time $\tau$.

For bright targets, the overheads for small apertures are dominated by the time required to dig the initial dark hole, while the overheads for large apertures are dominated by the slew time.  For faint targets, the overheads are dominated by the dark hole touch-up for both large and small apertures.

\subsection{Starshade mission parameters}

For starshade-based missions, we made as similar of assumptions as possible.  We assumed the same raw contrast floor, noise floor, reflectivities/transmissivities, and detector.  Because starshades have relatively high core throughput, and because their yields are not directly limited by exposure time\cite{stark2016_2}, we assumed starshade-based missions would have access to the UV and considered only a single low-throughput scenario.  Optical throughputs for the starshade instruments were estimated using a simplified instrument layout and informed by the HabEx A starshade instrument design\cite{habex_interim}. Table \ref{SS_params_table} summarizes the parameters we adopted for the starshade-based missions.

Table \ref{SS_params_table} also lists the propulsion and mass assumptions for the starshade. These parameters are critical to estimating yields, as starshade-based missions are typically fuel-limited. All propulsion and mass estimates come from current HabEx A starshade designs for a 4 m telescope\cite{habex_interim}. We assumed that starshade mass scales linearly with starshade diameter\cite{stark2016}, that fuel mass is equal to dry mass to ensure efficient propulsion, and that starshade separation is determined by the starshade diameter and the IWA. While it is technically true that the IWA of starshades is independent of telescope diameter, in practice one would not work interior to the diffraction limit, as the image would become unresolved and planets in the system would be indistinguishable; thus, we set the IWA of the starshade to $1.2$$\lambda/D$ at the longest wavelength of the starshade's nominal bandpass, 1 $\mu$m, such that the starshade IWA scales as $D^{-1}$.

Because starshades block starlight before it reaches the telescope, they do not need ultra-stable optics.  As a result, the settling time is expected to be much less than that for coronagraph-based missions; we adopt 10 minutes for the average slew and settle time. However, starshades do require significant overheads to precisely align the starshade with the telescope, estimated at 6 hours\cite{habex_interim}.

As discussed in detail in Section \ref{obs_strategy_section}, we adopt an observing strategy designed to play to the strengths of the starshade and maximize its yield. This strategy requires minimizing fuel use/slews, a different approach from that of the coronagraph-based missions. Thus, to differentiate between planets and obtain as much information as possible in a single visit, instead of simple broadband detections we require initial $R=70$, S/N $=5$ spectra for every observation using an IFS. This, combined with the assumed presence of a UV starshade camera, results in the relatively low optical throughput for exoplanet detections listed in Table \ref{SS_params_table}. Because starshades have wide bandpasses ($\sim$100\%) and high core throughput at the IWA, they are capable of producing high quality ``complete" spectra in a single observation; thus we take advantage of this and assume all exoEarths are followed up with $R=140$, S/N $=10$ spectra.

\begin{deluxetable}{ccl}
\tablewidth{0pt}
\footnotesize
\tablecaption{Starshade-based Mission Parameters\label{SS_params_table}}
\tablehead{
\colhead{Parameter} & \colhead{Value} & \colhead{Description} \\
}
\startdata
& & \bf{General Parameters} \\
$\Sigma \tau$ & $\sim2$ yrs & Total exoplanet science time of the mission\tablenotemark{*} \\
$\tau_{\rm slew}$ & 10 min & Static overhead for slew and settling time \\
$\tau_{\rm align}$ & 6 hrs & Static overhead to align starshade with telescope\\
$X$ & $0.7$ & Photometric aperture radius in $\lambda/D$ \\
$\Omega$ & $\pi(X\lambda/D)^2$ radians & Solid angle subtended by photometric aperture \\
$\zeta_{\rm floor}$ & $10^{-10}$ & Raw contrast floor \\
$\Delta$mag$_{\rm floor}$ & $26.5$ & Noise floor (faintest detectable point source at S/N$_{\rm d}$) \\
$T_{\rm contam}$ & $0.95$ & Effective throughput due to contamination \\
$\lambda_{\rm min}$ & $0.3$ $\mu$m & Minimum wavelength of nominal starshade bandpass \\
$\lambda_{\rm max}$ & $1.0$ $\mu$m & Maximum wavelength of nominal starshade bandpass \\
$D_{\rm SS}$ & 52 m & Starshade diameter\tablenotemark{**} \\
$z_{\rm SS}$ & $76.6$ Mm & Starshade-telescope separation distance\tablenotemark{**} \\
IWA & $1.2\lambda_{\rm max}/D$ & Starshade inner working angle \\

\hline
& & \bf{Detection Parameters} \\
$\lambda_{\rm d}$ & $0.65$ $\mu$m & Central wavelength for detection \\
S/N$_{\rm d}$ & $5$ & S/N required for detection (per spectral resolution element) \\
$R_{\rm d}$ & $70$ & Spectral resolving power required for detection \\
$T_{\rm optical,IFS}$ & $0.35$ & End-to-end reflectivity/transmissivity at $\lambda_{\rm d}$ \\
$\tau_{\rm d,limit}$ & 2 mos & Detection time limit including overheads \\

\hline
& & \bf{Characterization Parameters} \\
$\lambda_{\rm c}$ & $0.65$ $\mu$m & Wavelength for characterization \\
S/N$_{\rm c}$ & $10$ & Signal to noise per spectral bin evaluated in continuum \\
R & 140 & Spectral resolving power \\
$T_{\rm optical,IFS}$ & $0.35$ & End-to-end reflectivity/transmissivity at $\lambda_{\rm c}$ \\
$\tau_{\rm c,limit}$ & 2 mos & Characterization time limit including overheads \\

\hline
& & \bf{Detector Parameters} \\
$n_{\rm pix,d}$ & 72 & \# of pixels per spectral bin in IFS at $\lambda_{\rm d}$\\
$n_{\rm pix,c}$ & 72 & \# of pixels per spectral bin in IFS at $\lambda_{\rm c}$ \\
$\xi$ & $3\times10^{-5}$ $e^-$ pix$^{-1}$ s$^{-1}$ & Dark current\\
RN & 0 $e^-$ pix$^{-1}$ read$^{-1}$ & Read noise\\
$\tau_{\rm read}$& N/A & Time between reads\\
CIC & $1.3\times10^{-3}$ $e^-$ pix$^{-1}$ frame$^{-1}$ & Clock induced charge\\
$T_{\rm QE}$ & $0.9$ & Raw QE of the detector at all wavelengths \\
$T_{\rm read}$ & $0.75$ & Effective throughput due to bad pixel/cosmic ray mitigation \\

\hline
& & \bf{Propulsion Parameters} \\
$m_{\rm dry}$ & 4550 kg & Starshade dry mass\tablenotemark{**} \\
$m_{\rm fuel}$ & $m_{\rm dry}$ & Fuel mass \\
$I_{\rm sk}$ & $300$ s & Specific impulse of station keeping propellant (chemical)\\
$I_{\rm slew}$ & $3000$ s & Specific impulse of slew propellant (electric)\\
$\epsilon_{\rm sk}$ & $0.8$ & Efficiency of station keeping fuel use\\
$\epsilon_{\rm slew}$ & $0.9$ & Efficiency of slew fuel use\tablenotemark{***}\\
$\mathcal{T}$ & $1.04$ N & Thrust for slewing\\

\hline
\enddata
\vspace{-0.1in}
\tablenotetext{*}{Exposure time is optimally balanced with slew time\cite{stark2016_2}.}
\tablenotetext{**}{For 4 m telescope; scales with telescope diameter.}
\tablenotetext{***}{Value optimized for each simulation to maximize yield\cite{stark2016_2}.}

\end{deluxetable}

\section{Telescope Scenarios and Instrument Design}

\subsection{Coronagraph Design\label{coronagraph_section}}

Perhaps more than any other type of instrument, coronagraphs must be designed with the telescope and spacecraft as a system.  Seemingly minor decisions about, e.g., the geometry of the outer edge of the primary mirror or the size of gaps between segments, can dramatically impact performance.  Larger decisions about the primary mirror (PM) and secondary mirror (SM) geometry can completely rule out certain classes of coronagraphs.  Our goal is to understand how these decisions ultimately impact exoplanet yield of the mission.  

In addition to the low and high throughput scenarios described above, we considered two major design decisions: on- vs off-axis secondary mirrors, and monolithic vs segmented primary mirrors.  Given that off-axis telescopes are generally more feasible at smaller apertures, we studied these two major decisions by considering three critical OTA scenarios: off-axis monolithic telescopes ($D\lesssim4.5$ m), off-axis segmented telescopes ($4.5\lesssim D \lesssim 9$ m), and on-axis segmented telescopes ($D \gtrsim 4.5$ m).  Here we use the term ``on-axis" loosely, to denote that the SM creates an obscuration centered on the PM, though technically the optical axis could still be slightly off-axis.

\begin{figure}[H]
\centering
\includegraphics[width=6.5in,trim={3.1in 1.1in 4.6in 1.9in},clip]{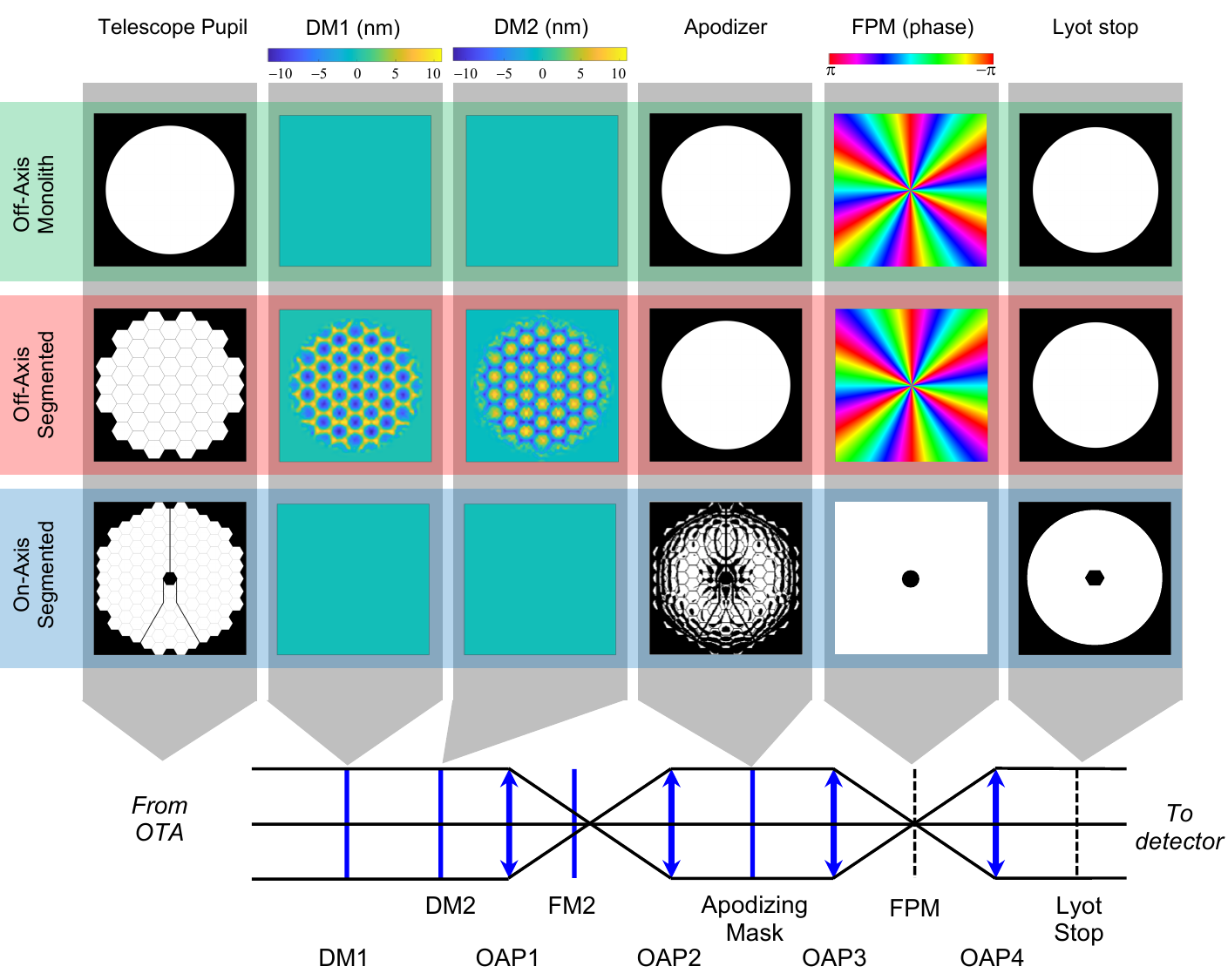}
\caption{The optical layout for the adopted coronagraphs in each of the three OTA scenarios. All black and white masks are binary apodization masks.  The off-axis scenarios use a phase-shifting focal plane mask, while the on-axis scenario uses an apodized focal plane mask.  \label{coronagraphs_fig}}
\end{figure}

For each of these scenarios, we necessarily assumed a PM-SM geometry.  For the off-axis monolith, the aperture is a simple unocculted disk.  For segmented scenarios, we followed these general rules of thumb that lead to improved coronagraph performance:
\begin{itemize}
	\item {\bf Maximize the inscribed diameter}---coronagraphs do not appear to efficiently use the jagged outer edge of a hex-patterned segmented primary. Broadband vortex coronagraphs (VC) can currently handle only circular or elliptical pupils\cite{ruane13}, and thus are strictly limited to the sub-aperture defined by the inscribed primary diameter\cite{ruane2018}. Optimized apodized pupil Lyot coronagraph (APLC) pupil masks tend to largely discard the primary beyond the inscribed diameter\cite{stlaurent2018}.
	\item {\bf Minimize the secondary obscuration}---large secondary obscurations penalize  performance in two ways.  First, large secondary obscurations produce a telescope PSF that is significantly broader than an Airy pattern, such that there is less light in the PSF core.  Second, they reduce the throughput of APLCs\cite{stlaurent2018} and lead to extreme stellar diameter sensitivity with VCs\cite{ruane2018}.
	\item {\bf Minimize the segment gap width}---as long as the gap width is small, diffraction due to the gaps can be corrected using the deformable mirrors (DM).  Segment gaps become less of a problem as the primary increases in size, as it is the gap size relative to the inscribed diameter that matters\cite{ruane2018}.  We assumed $\sim$6 mm gaps, which equates to $0.1\%$ of the smallest inscribed diameter considered.  
\end{itemize}
With these rules of thumb in mind, we adopted the primary mirror geometries shown on the left in Figure \ref{coronagraphs_fig}.

For the above OTA scenarios we considered the following pool of coronagraphs, as summarized in Refs.~\citenum{ruane2018} and \citenum{guyon2014}: APLCs, DM-assisted APLCS (DMAPLC), VCs, Apodized Vortex Coronagraphs (AVC), DM-assisted Vortex Coronagraphs (DMVC), Hybrid Lyot Coronagraphs (HLC), and Phase Induced Amplitude Apodization Complex Mask Coronagraphs (PIAACMC).  All of these coronagraphs use the same basic optical layout, but adopt different optics. Figure \ref{coronagraphs_fig} illustrates the optical layout of a single coronagraph (a subset of Figure \ref{optical_layout_fig}) along with the optics used for three of these coronagraph designs.  We numerically simulated the optical performance of these coronagraph designs. Details of the numerical design methods can be found in Refs.~\citenum{ruane2018} and \citenum{stlaurent2018}, as well as references therein.

For each coronagraph design we produced two primary data products following a standardized format for input to our yield calculations \cite{starkkrist}.  First, we calculated the on-axis leaked starlight normalized to the flux entering the coronagraph, $I'(x,y,\theta_{\star})$.  This leaked starlight degrades as stellar diameter increases, and different coronagraph designs can have drastically different sensitivities to stellar diameter. To include these effects in our calculations, we generated the 3D data cube $I'(x,y,\theta_{\star})$ with pixel size $\theta$, where $x$ and $y$ are on-sky coordinates relative to the center of the coronagraph and $\theta_{\star}$ is stellar diameter. Both $\theta$ and $\theta_{\star}$ are in units of $\lambda/D$, allowing us to apply these simulations at any wavelength and telescope diameter.  We typically evaluated $I'(x,y,\theta_{\star})$ for roughly one dozen stellar diameters, ranging from 0--10 mas for the expected range of $D$, with greater sampling near $\sim1$ mas.  The sampling was chosen independently for each coronagraph design to resolve changes in the raw contrast as a function of stellar diameter, enabling accurate interpolation.

To calculate exposure times using Eq. \ref{CRbstar_equation} for a given star of diameter $\theta_{\star}'$ and a given planet at $\left(x,y\right)$, we interpolated this 3D grid to the desired stellar diameter value, $I \!\left(x,y\right)= I'\!\left(x,y,\theta_{\star}\!=\!\theta_{\star}'\right)$, i.e. coronagraph contrast is calculated individually for each star based on stellar diameter.  

Second, we simulated a set of off-axis PSFs normalized to the flux incident on the coronagraph, $P(x,y,\Delta x,\Delta y)$, where $\Delta x$ and $\Delta y$ are the offsets of the source with respect to the center of the coronagraph.  Using these off-axis PSFs and an assumed photometric aperture solid angle $\Omega$, we calculated $\Upsilon_{\rm c}\!\left(x,y\right)$.

Alignment, jitter, stability, and wavefront sensing and control (WFSC) may have significant impacts on the  coronagraph performance.  However, realistic simulations of these factors require a detailed engineering model of the full system, something that is beyond the scope of this paper.  In lieu of this, we made a few simple assumptions that avoid overly optimistic performance.  First, the response of a coronagraph to jitter is similar to the response to stellar diameter, so we simply assumed that the RMS jitter is less than the stellar diameter.  Second, we assumed that the coronagraph can achieve a static raw contrast no better than $\zeta_{\rm floor} = 10^{-10}$; we substituted a contrast of $10^{-10}$ wherever $I\!\left(x,y\right)$ exceeded this performance.  We note that this contrast limit does not set the post-processing noise floor, which we assumed is consistent with detecting a planet with $\Delta {\rm mag_p}=26.5$ (corresponding to a post-processing factor of 10--20 improvement depending on the assumed S/N).

\subsubsection{Coronagraph selection optimization}

Because the optical layouts of most coronagraphs are fairly similar, future missions will likely carry multiple types of coronagraphs on board and swap between them by rotating several pupil plane and/or image plane wheels.  With this in mind, the yield for a given telescope scenario may be maximized by mixing together several types of coronagraphs.

To determine the optimum set of coronagraphs for each OTA scenario, we modified our yield code to handle up to four coronagraphs for detection and four (potentially different) coronagraphs for spectral characterization.  Ideally we would optimize the selected coronagraph on an observation-by-observation basis, based on the observation's benefit:cost ratio within the main Altruistic Yield Optimization (AYO) function\cite{stark2014_2}.  However, this would dramatically increase the complexity and run time of the code.  As a simpler alternative, we assigned a single coronagraph to each star based on a cursory benefit:cost ratio analysis.  For each star, we calculated the first visit completeness as a function of exposure time for each coronagraph, determined the peak of the completeness divided by exposure time, and choose the coronagraph with the largest peak value.

In practice, this usually selected a single coronagraph design for all stars for each OTA scenario.  The exception is the segmented on-axis scenario, which selects multiple APLC coronagraphs.  For brevity, we present below only the best performance coronagraph(s) for each OTA scenario, along with their optical layouts in Figure \ref{coronagraphs_fig}.  We note that as coronagraph designs continue to improve, the best performance coronagraph may change; the performance and yield presented herein should be thought of as a snapshot in time.

\subsubsection{Monolithic off-axis telescope: VC}

For the monolithic off-axis OTA scenario, the best performance coronagraph design is a VC.  A VC \cite{mawet2005,foo2005} consists of a phase-only focal plane mask whose spatially-dependent transmission can be expressed as $e^{ilq}$, where $l$ is an integer known as the charge and $q$ is the azimuthal angle (see Figure \ref{coronagraphs_fig}). For even, nonzero values of $l$ and a circular, unobstructed entrance aperture, all of the light from a point source on the optical axis is relocated outside of the geometric image of the following pupil. A Lyot stop whose radius is less than the geometric pupil radius then blocks all of the starlight, preventing it from reaching the final image plane. 

The best coronagraph contrast measured to date was reported in Ref.~\citenum{serabyn2013} and obtained on the JPL high contrast imaging testbed (HCIT) with a second-generation charge 4 vector vortex coronagraph made out of liquid crystal polymers (LCP) by JDS Uniphase. Monochromatic tests using a laser diode source at a wavelength of 785 nm produced an all-time best raw contrast result of $4.1\times10^{-10}$ in a 3--8 $\lambda/D$ half-disk. Broadband tests using a supercontinuum laser source and a series of five adjacent 2\%-bandwidth spectral filters (net bandwidth of 10\%) yielded an average suppression of $5.0\times10^{-9}$ when ignoring a bright speckle due to contamination or a manufacturing defect. An on-going Technology Demonstration for Exoplanet Missions (TDEM) program led by E. Serabyn (JPL) is aimed at testing third-generation achromatic LCP vector vortex masks with topological charges of 4--8 with a goal of demonstrating $10^{-9}$ raw contrast over 10\% and 20\% bandwidths.

While lower charge VCs provide higher planet throughput at smaller angular separations, higher charge VCs reduce the amount of leaked starlight due to tip/tilt errors and large stellar diameters. Higher charge also provides a means to relax wavefront error requirements; a VC has relaxed wavefront error requirements ($>100$ pm RMS) for $l^2/4$ Zernike aberrations. Denoting Zernike aberrations by $Z_n^m$, where $n$ and $m$ are the radial and azimuthal indices, these are modes where $|l| > n+|m|$. For all other low order aberrations, the wavefront error requirements are $\sim1$ pm RMS \cite{ruane2018b}.  Given the typical size of nearby stars and the expected levels of aberration in the telescope, we expect our adopted charge 6 VC to provide the best performance. 

\begin{figure}[H]
\centering
\includegraphics[width=4.5in]{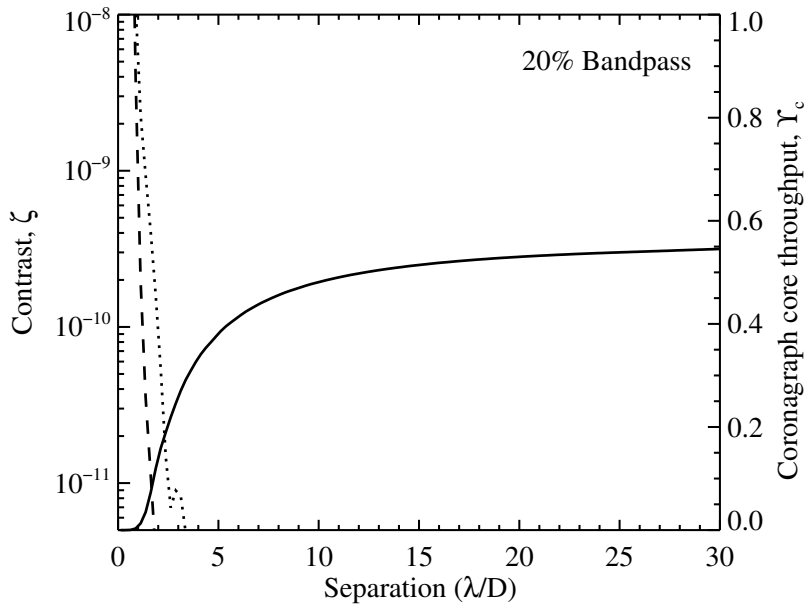}
\caption{Azimuthally-averaged raw contrast $\zeta$ as a function of separation for an on-axis point source (dashed) and an on-axis source with diameter $0.1\lambda/D$ (dotted) for the VC charge 6 with an off-axis monolithic OTA. During yield calculations, we set the contrast to the greater of $\zeta$ and $\zeta_{\rm floor}$.  Adopted core throughput $\Upsilon_{\rm c}$ is also shown (solid).   \label{VC_fig}}
\end{figure}

Figure \ref{VC_fig} shows the VC's azimuthally averaged raw contrast for a point source (dashed line) assuming perfect optics.  A VC with an unperturbed off-axis monolithic primary produces a perfect null at large separations.  Of course in practice, this will not be the case.  For our yield calculations, we adopted a contrast floor $\zeta_{\rm floor} = 10^{-10}$, such that the assumed raw contrast was never better than $10^{-10}$. Figure \ref{VC_fig} also shows the raw contrast for a star with diameter $0.1\lambda/D$ (dotted line). The most common observed stellar diameter will be $\sim0.6$ mas, equating to $\sim0.03\lambda/D$ for a 4 m telescope at V band; the raw contrast should typically be between the dashed and dotted curves in Figure \ref{VC_fig} when applied to individual stars.

Figure \ref{VC_fig} also shows the azimuthally averaged coronagraphic core throughput $\Upsilon_{\rm c}$ for off-axis PSFs (solid line).  As this curve shows, a planet's throughput is a slowly increasing function of separation.  Traditionally coronagraphs are parameterized in terms of contrast, throughput, bandwidth, outer working angle (OWA), and inner working angle (IWA).  IWA is typically defined as the separation at which the off-axis PSF throughput reaches half of its maximum value, in this case $\sim$3 $\lambda/D$.  Oftentimes the IWA is interpreted to be a region on the sky interior to which planets cannot be detected.  However, as Figure \ref{VC_fig} shows, this interpretation of IWA is not entirely consistent with the definition; useful throughput exists down to $\sim$1.5 $\lambda/D$, and the contrast is below $10^{-10}$ at similar separations.  I.e., VCs can detect planets inside of their classically-defined IWA, as long as one can pay the throughput penalty.  We note that our AYO code automatically determines whether to do so when it uses the $\Upsilon_{\rm c}$ curve to optimize target selection and exposure times.

\subsubsection{Segmented off-axis telescope: DMVC}

For off-axis segmented apertures, VCs may be designed to achieve similar performance characteristics as monolithic apertures via apodization techniques \cite{ruane2018}. Unwanted diffraction from gaps between mirror segments can be controlled either actively with two DMs (DMVCs) or by using static pupil masks with grayscale transmission (AVCs). The former provides higher throughput, but requires gaps sizes  $\lesssim0.1\%$ of the full pupil diameter. The latter can be used to achieve comparable raw contrast with larger gaps at the cost of throughput. Overall, smaller gap sizes lead to better theoretical performance in terms of throughput as well as raw contrast across a given spectral bandwidth. For the off-axis segmented OTA scenario, we adopted a DMVC.

\begin{figure}[H]
\centering
\includegraphics[width=4.5in]{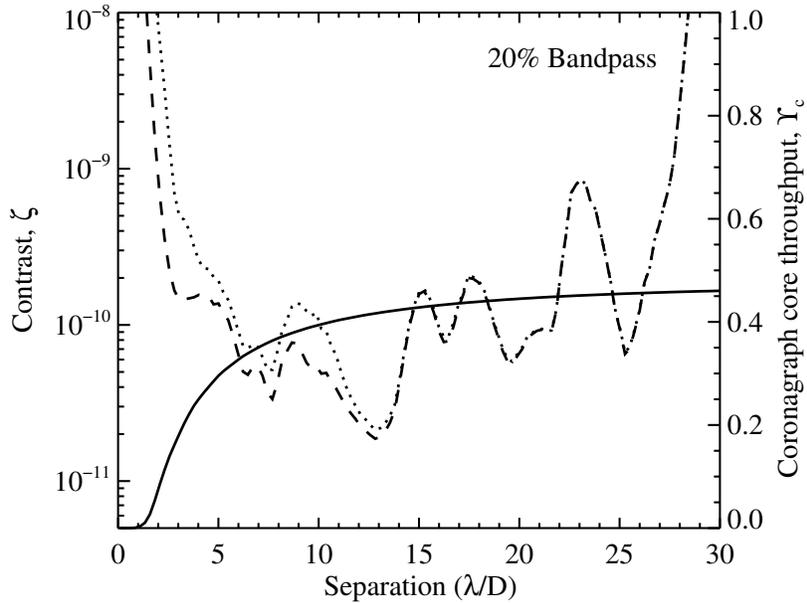}
\caption{Same as Fig.~\ref{VC_fig}, but for the DMVC with a segmented off-axis OTA.  The adopted core throughput appears smaller for the DMVC than the VC because it is normalized to the full obscured primary (including the pupil exterior to the inscribed diameter that is discarded by the Lyot stop); normalizing to the inscribed diameter pupil would make throughputs appear nearly equal.    \label{DMVC_fig}}
\end{figure}

The DMVC contrast and throughput curves are shown in Figure \ref{DMVC_fig}. Because the DMs cannot perfectly correct for the gaps in the pupil, the contrast at large separations is not infinitely good, unlike the VC for a monolithic aperture.  For a $\sim$7 m inscribed diameter telescope at V band, a typical star diameter is $\sim0.04 \lambda/D$; the contrast should commonly be between the dashed and dotted lines.

Although $\Upsilon_{\rm c}$ for the DMVC appears to be significantly less than the VC, it is due simply to the normalization.  $\Upsilon_{\rm c}$ is the fraction of the light exiting the corongraph within the photometric aperture normalized to the light entering the coronagraph.  The light entering the coronagraph comes from the full obscured primary mirror, including the region exterior to the inscribed diameter (the jagged outer edge of the primary).  Because the DMVC can only handle circular or elliptical pupils, it is therefore limited to the inscribed diameter and the Lyot stop must discard the light from the outer edges of the segmented primary.  This discarded light from beyond the inscribed diameter is what causes the lower apparent core throughput.  However, if we were to plot the core throughput of the VC and DMVC normalized to the inscribed pupil, they would look nearly identical.

\subsubsection{Segmented on-axis telescope: Multi-mask APLC}

For the segmented on-axis OTA scenario, the best performance coronagraph was a suite of APLC designs.  Recent APLC designs have focused on telescope apertures obscured by secondary mirrors, struts, and segment gaps \cite{ndiaye2016,zimmerman2016a}. These designs utilize a binary-valued shaped pupil apodizer to filter the obscuration features of the telescope pupil, mimicking the Fourier properties of a circular prolate spheroidal wavefunction \cite{soummer2003}, in addition to hard-edged, opaque occulting masks.  These designs generally have less favorable IWA than coronagraphs that modify the phase in the occulting plane (e.g., the VC). Since the apodizer mask reduces the throughput of the off-axis exoplanet PSF, the typical design strategy is to maximize the apodizer transmission for a given set of contrast and working angle constraints \cite{carlotti2011}. Shaped pupil apodizer masks have a relatively mature fabrication process, and laboratory prototypes for the WFIRST Coronagraph Instrument have successfully demonstrated broadband dark search zones at $\sim10^{-9}$ contrast \cite{cady2017,marx2018}.

The ExEP's Segmented Coronagraph Design and Analysis study has supported large design parameter surveys of APLC designs for centrally obscured, segmented pupils \cite{zimmerman2016b,stlaurent2018}. From these investigations, a few general properties have emerged:
\begin{itemize}
\item APLC designs for centrally obscured apertures encounter sharp lower limits in IWA; for a given bandpass and contrast goal, the apodizer throughput drops sharply when the occulter radius is reduced below a certain threshold, while for larger occulter radii the throughput plateaus. The most aggressive APLC designs for LUVOIR-like apertures achieve an IWA of $\sim$3.5 $\lambda/D$.
\item Only relatively small modifications to the apodization pattern are needed to accommodate struts and segment gaps (generally 1\% or less of pupil diameter). The dependence of apodizer throughput on the orientations of the struts and segment gaps is marginal.
\item APLC designs can adapt to relatively large central obstructions. The obscuration can be increased up to $\sim$25\% of diameter before the apodizer transmission from the illuminated region of primary mirror sharply degrades. However, design solutions for large central obscurations (e.g., WFIRST at 31\%) generally require larger occulting masks, resulting in poor IWA.
\item When the occulting mask has a large outer edge (field stop), the PSF is highly sensitive to Lyot stop alignment. This can be mitigated by deformable mirror control and compound optimization constraints incorporating propagation through misaligned masks.
\item The APLC is intrinsically robust to stellar diameters up to $\sim0.1 \lambda/D$. For reference, $0.1 \lambda/D$ corresponds to approximately 1 mas at visible wavelengths for a 12-meter diameter primary mirror, or equivalently, the angular diameter of a solar twin at a distance of 10 pc.
\item The apodizer throughput depends on the ratio of the OWA and IWA. For OWA/IWA $\gtrsim3$, the throughput tapers off. As a result, multiple sets of masks (apodizers and occulting masks) are needed for an instrument to cover a wide range in working angles.
\end{itemize}

\begin{figure}[H]
\centering
\includegraphics[width=4.5in]{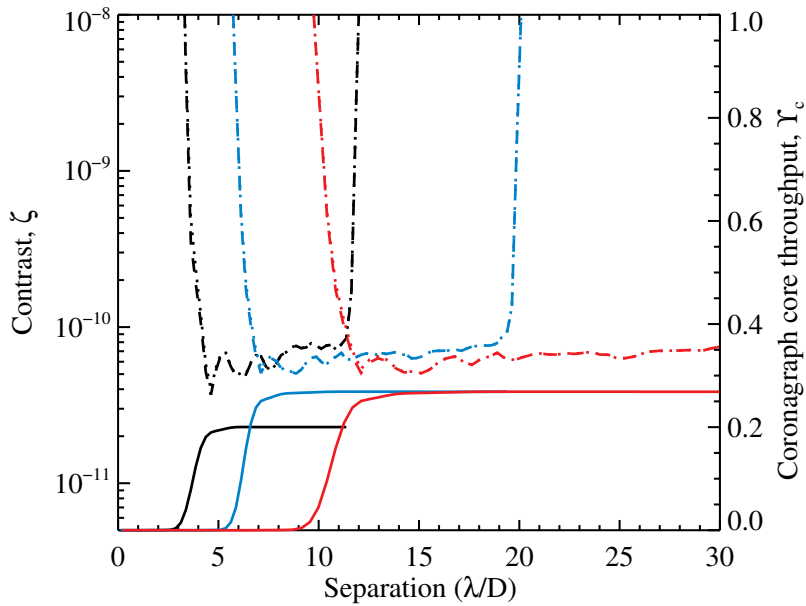}
\caption{Same as Fig.~\ref{VC_fig}, but for the APLC with a segmented on-axis OTA.  The results for all three adopted masks are shown in black, blue, and red.  The APLC is very robust to stellar diameter; the dotted and dashed lines are nearly coincident.    \label{APLC_fig}}
\end{figure}

To maintain both high throughput and a field of view that covers a broad range of HZ sizes, we set OWA/IWA $\sim3$ and developed a suite of 3 masks with overlapping dark zones, as shown in Figure \ref{APLC_fig}.  The overlap of these masks was chosen to ensure that any HZ observed would have a corresponding mask that could image the inner and outer edges of the HZ simultaneously.  Details of this overlap optimization are covered in Ref.~\citenum{stlaurent2018}. We designed a small-HZ mask covering $\sim3.6-11\lambda/D$, a medium-HZ mask covering $\sim6-19\lambda/D$, and a large-HZ mask covering $\sim11-33\lambda/D$.  In practice, the numerical design of large-OWA masks is computationally expensive.  Because of this, we took a reasonable and conservative shortcut for our large-HZ mask design---we simply assumed its performance was equivalent to the medium-HZ mask, but with increased pixel scale.

Because the SM obscuration ratio decreases as telescope diameter increases,  we created designs for two bounding values of SM obscuration: 10\% and 30\%.  We interpolated our results to calculate the performance for intermediate SM obscuration ratios.

Figure \ref{APLC_fig} shows the contrast and throughput curves for the three APLC masks adopted (black, blue, and red curves).  The dashed and dotted lines are nearly coincident because the APLC is insensitive to stellar diameter, a very beneficial property given the larger telescopes to which this design is applied.  Unlike the VC and DMVC, the throughput curves closely resemble top-hat functions and the contrast is relatively uniform over the dark hole; the typical notions of IWA apply to the APLC much better than the VC.

\subsection{Starshade Design\label{coronagraph_section}}

\begin{figure}[H]
\centering
\includegraphics[width=4.5in]{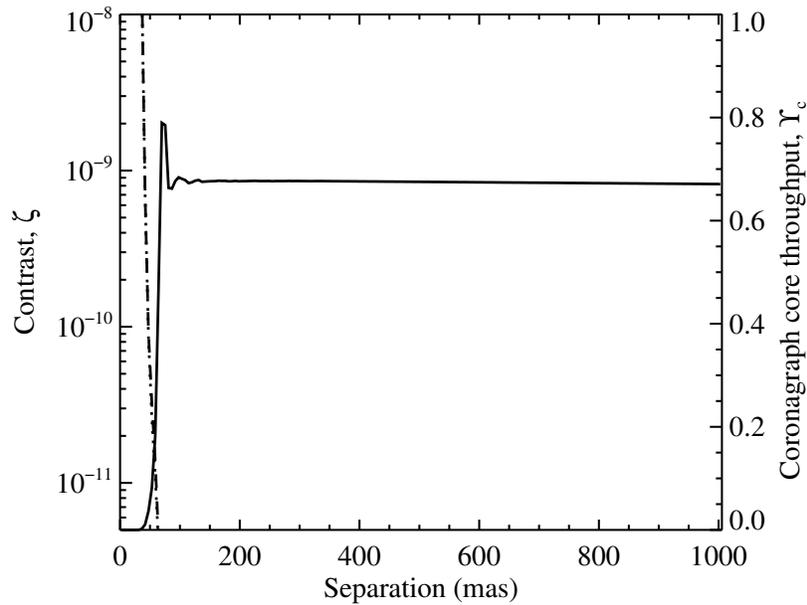}
\caption{Same as Fig.~\ref{VC_fig}, but for the baseline starshade paired with a 4 m telescope, valid from $0.3$--$1.0$ $\mu$m.\label{TV3_fig}}
\end{figure}

For our baseline starshade, we adopted the same design as for HabEx A: a 52 m diameter starshade separated by 76.6 Mm from the telescope, with an IWA of 60 mas, where the IWA is defined as the angle at which the core throughput falls to half its maximum value.  We note that while we use the baseline starshade design from HabEx A for our starshade-only calculations, HabEx A is a hybrid design using both a coronagraph and a starshade; we do not model the yields for this mission concept here. The starshade has a nominal bandpass of $0.3$-$1.0$ $\mu$m, such that the IWA is set to $1.2\lambda/D$ at 1 $\mu$m.  We simulated the optical performance of the starshade using a wave propagation model and created inputs for the yield code using the same standards developed for coronagraphs above. Figure \ref{TV3_fig} shows the performance of the starshade; a raw contrast floor of $10^{-10}$ was enforced at all separations.

For a fixed IWA, starshade diameter depends only weakly on telescope diameter\cite{stark2016_2}. We made the approximation that starshade diameter has no direct dependance on $D$, but chose to tie the starshade IWA (and thus diameter) to the telescope's diffraction limit.  Thus, to model the optical performance of starshade-based missions with varying telescope diameters, we simply adjusted the pixel scale of the baseline starshade simulations by the telescope's diameter.

\section{Adopted observation strategies\label{obs_strategy_section}}

Coronagraphs and starshades differ significantly in design, performance, and implementation.  Efficient use of these two instruments requires different observation strategies, and as a result they would produce different data sets. We adopted the observing strategies suggested by Ref.~\citenum{stark2016} that simultaneously maximize yield while providing measured colors, orbits, phase variations, and spectra for all exoEarth candidates detected.

For coronagraph-based missions, we assumed that following initial broadband detection, planets would be differentiated via measurement of their color, phase variations, and orbits.  We required at least 6 visits to each star to account for orbit determination and multi-epoch photometry.  Recent work by Ref.~\citenum{guimond2019} suggests that as few as three visits may be necessary to constrain a planet to within the habitable zone assuming 5 mas astrometric precision and a visit cadence $>1/4$ orbit.  However, this result relies on using both the presence and absence of planets in observations to constrain the orbit, but does not include the effects of a noise floor, which may make it difficult to determine whether a crescent-phase planet is absent.  In light of this, we erred on the side of caution and continued to adopt 6 visits as a reasonable number.  For the coronagraph-based missions, the exposure times of each of the six visits are not equal; they were optimized using AYO to maximize yield.

After orbit-determination, for coronagraph-based missions we required that all exoEarth candidates be followed up with cursory spectral characterization to search for water vapor at 950 nm, estimated as a single $R$=70, S/N$_{\rm c}$=5 spectrum, evaluated at 1 $\mu$m \cite{brandt2014, luvoir_interim}.  We assumed that the orbit was well-determined, such that the phase of the planet could be optimized.  Spectral characterization times were probabilistically calculated using the methods of Ref.~\citenum{stark2015}. We assumed that the multi-epoch color photometry and orbit determination allowed us to perfectly differentiate exoEarth candidates from all other planets, an assumption that may be a bit optimistic\cite{guimond2018}.  A reduced ability to differentiate planets would decrease yields \cite{stark2016}.

For starshades, we adopted a different observing strategy to minimize fuel use \cite{stark2016}.  The adopted scenario requires initial cursory $R$=70, S/N$_{\rm c}$=5 spectra on every planetary system to detect and identify exoEarth candidates in a single observation. Only exoEarth candidates are then followed up with an $R$=140, S/N$_{\rm c}$=10 spectrum (to detect O$_2$ and other key biosignatures) and 5 additional visits to measure orbits. Because starshades will obtain spectra prior to precisely measuring orbits, we did not allow the phase of the planets to be optimized.

\section{Results \& Discussion}
\label{results_section}

\begin{figure}[H]
\centering
\includegraphics[width=6.5in]{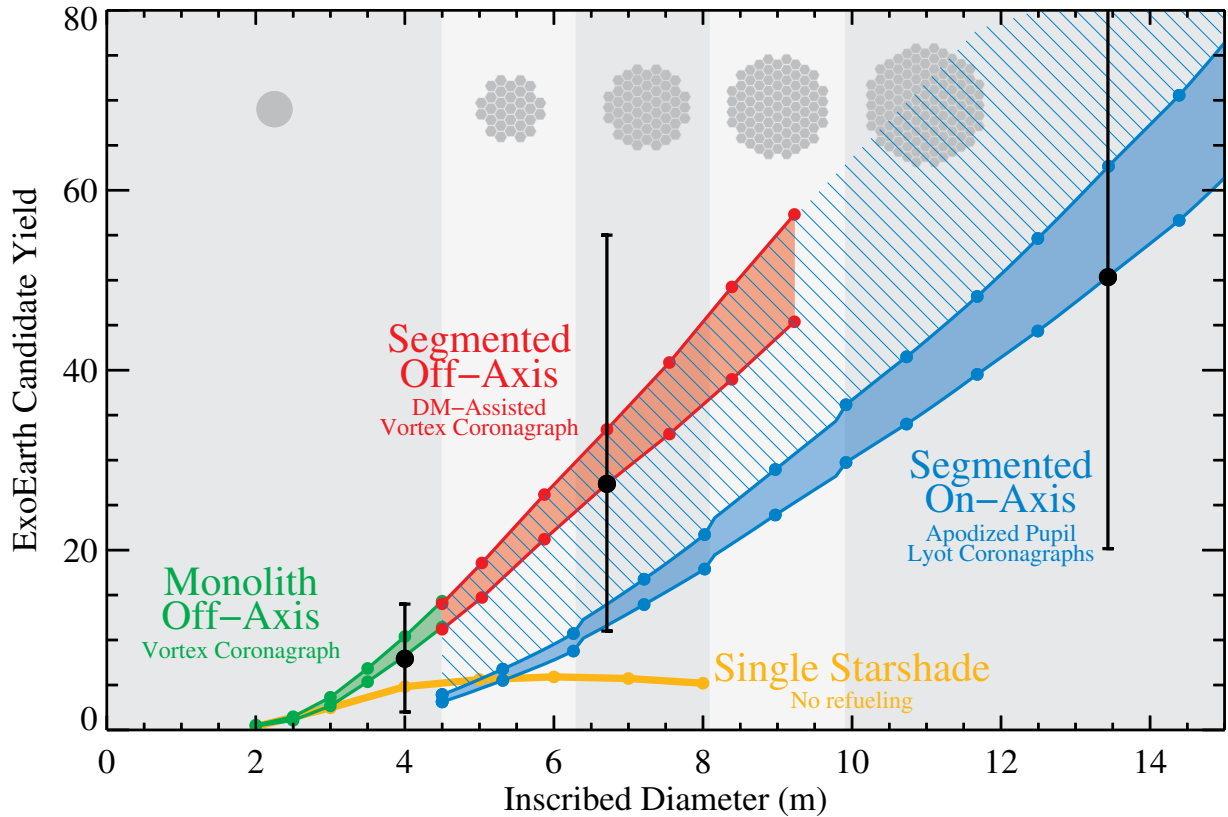}
\caption{The exoEarth candidate yield landscape for future direct imaging coronagraph- and starshade-based space telescopes, assuming 2 years of telescope time. The spread in coronagraph yields (green, red, and blue) corresponds to high and low throughput scenarios considered herein. Astrophysical uncertainties are shown for several point designs (black). Without refueling, the yields of starshade-based missions are limited, while yield increases with aperture size for coronagraph-based missions. As long as segment gaps are small and the requisite engineering constraints can be met, there is little to no yield penalty for segmentation, but currently a substantial penalty for an on-axis telescope design. The primary mirror geometry is illustrated along the top for each region of the plot; on-axis pupils would have the central segment removed. The blue hashed region illustrates the yield that may be possible with on-axis telescopes if the performance of future on-axis coronagraph designs improves.\label{yield_figure}}
\end{figure}

Using the above assumptions, we calculated the expected exoEarth candidate yield for all scenarios considered.  The green, red, and blue curves in Figure \ref{yield_figure} show our results for coronagraph-based missions for all OTA scenarios.  In each scenario, the lower and upper curves correspond to the low and high throughput scenarios, respectively.  An illustration of the primary mirror geometry is shown along the top of the plot, although it does not show the struts or SM for the on-axis scenario.

As Fig. \ref{yield_figure} shows, our simulations suggest there is no yield penalty for transitioning from a monolithic off-axis telescope to a segmented off-axis telescope, provided that segment gaps can be made small enough to be filled by the DMs.  In reality, it's possible that a small yield penalty exists due to our inability to manufacture gaps small enough, though this effect likely goes away at larger inscribed diameters.  Our simulations also largely ignored any additional stability and WFSC challenges that segmented mirrors may create; we simply assumed the engineering tolerances associated with a raw contrast of $10^{-10}$ could be met.

The distance between the PM and SM is much larger for an off-axis telescope than an on-axis telescope of equivalent diameter and maximum angle-of-incidence limitation. At some point this distance becomes prohibitively long, either due to packaging, deployment concerns, or SM stability, and we must transition to an on-axis telescope.  This likely occurs somewhere near an inscribed diameter of $\sim$9 m.  Figure \ref{yield_figure} shows that there is currently a significant penalty for transitioning to an on-axis telescope, in spite of our efforts to adopt a PM-SM geometry amenable to coronagraphy and using several coronagraph masks optimized for the HZ.

The blue curve for the segmented on-axis OTA scenario shows several small ``wiggles." These wiggles occur at the transitions to larger numbers of segment rings, which are required to maintain reasonable segment sizes ($0.9$ to $1.3$ m).  The wiggles occur because we assume only the central segment is obscured by the SM, such that the SM obscuration ratio decreases as more rings are added, resulting in slightly better coronagraph performance. 

We note that there is nothing fundamentally limiting the performance of on-axis coronagraphs to what is shown in Figure \ref{yield_figure}.  Coronagraphs for on-axis telescopes with small SM obscurations can in theory achieve close to the same theoretical maximum performance as their off-axis counterparts\cite{guyon2006}. But no coronagraph design has yet been invented that can deal with the SM obscuration without significantly impacting the throughput, IWA, or contrast.  This may simply be due to the relative maturity of off-axis coronagraphs compared to on-axis designs.  Indeed, the performance of coronagraphs for on-axis segmented telescopes has increased dramatically in just the past few years\cite{ndiaye2016,ruane2018,stlaurent2018}. For this reason, we emphasize that the blue curve in Figure \ref{yield_figure} is a snapshot in time; the blue hashed region above the solid blue curves represents future performance that may be possible with additional research.

The orange curve in Figure \ref{yield_figure} shows the exoEarth candidate yields for starshade-based missions as a function of aperture.  As expected, the yields turn over as diameter increases, because the starshade diameter, mass, and separation distance also increase, all of which contribute to more costly slews.  Additionally, as the mission detects more exoEarth candidates, it must devote more slews to measuring the orbits of those planets; in the absence of refueling, the yields of starshade-based missions are self-limiting. We remind the reader that we do not consider hybrid missions in which both a coronagraph and starshade are used simultaneously.  Such missions would have similar yields to the coronagraph-only missions, but higher quality and quantity of spectra\cite{habex_interim}.

The colored curves shown in Figure \ref{yield_figure} illustrate the yield expected value. For three point designs, shown in black, we show the estimated 1$\sigma$ astrophysical uncertainties. The uncertainties shown include all major astrophysical sources: occurrence rate uncertainties, exozodi uncertainties, and the Poisson noise associated with the planets and exozodi randomly assigned to individual stars. We emphasize that these uncertainties are estimates, because the uncertainty on $\eta_{\Earth}$ and the exozodi distribution, and more specifically their probability distribution, are not precisely known.

To estimate astrophysical uncertainties, we assumed that the underlying occurrence rate and exozodi distribution probability distributions are Gaussian and divided the parameter space into a 3$\times$3 grid, defined by the pessimistic, nominal, and optimistic scenarios for occurrence rates and exozodi levels, centered on $-$1, 0, and $+$1 $\sigma$ in each dimension. For each of these 9 scenarios, we performed 20 yield calculations to sample the random assignment of exozodi levels to individual stars, for a total of 180 yield calculations.  Each of those 180 yield calculations produces an optimized observation plan with a detection probability for every observation.  Using these detection probabilities, we performed hundreds of Monte Carlo draws for each of the 180 yield calculations to sample the Poisson noise associated with the planetary systems of individual stars.  For each of the 9 scenarios, the number of Monte Carlo simulations was chosen to be proportional to the fraction of the 2D Gaussian probability distribution represented by each scenario.  We then calculated the 1$\sigma$ uncertainties in yield by sorting the thousands of Monte Carlo draws from smallest to largest, and finding the middle 68\% of yield values.  We note that our simplification of a 3$\times$3 grid, which covers only $-1.5\sigma$ to $+1.5\sigma$, misses 25\% of possible values in the tales of the Gaussian. However, the underlying assumption of Gaussian distributions is almost certainly incorrect; we consider this approximation as a reasonable estimate of astrophysical uncertainties given our current limited knowledge.

\begin{figure}[H]
\centering
\includegraphics[width=4in]{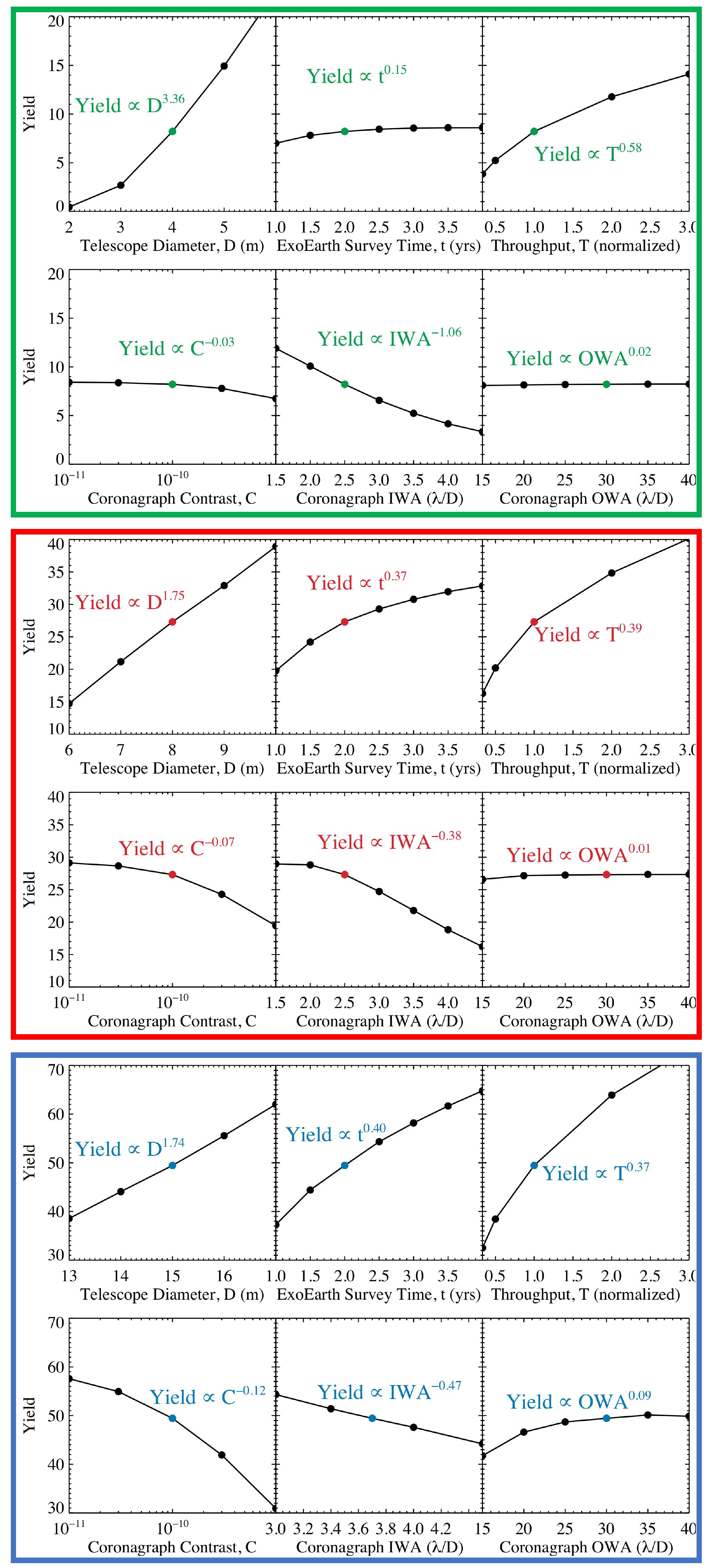}
\caption{The exoEarth candidate yield sensitivity to changes in one parameter at a time for each of the black points shown in Figure \ref{yield_figure}. Yield is a relatively weak function of most parameters, with the exception of telescope diameter.\label{sensitivity_figure}}
\end{figure}

Ref.~\citenum{stark2014_2} showed that yield is a relatively weak function of most mission/astrophysical parameters, with the exception of telescope aperture size, and also showed that the sensitivity varies depending on the parameters describing the mission.  Given the increased fidelity of these simulations and the different OTA scenarios examined, recalculation of these sensitivities is warranted. Figure \ref{sensitivity_figure} shows the response of the yield to changes in one mission/instrument parameter at a time for each of the three black points shown in Figure \ref{yield_figure}.

We confirm that yield is most sensitive to telescope diameter and broadly replicate the results of Ref.~\citenum{stark2014_2}. However, we note a few subtle differences.  First, the yield is an exceptionally strong function of diameter and weak function of total exposure time near $D=4$ m, likely a result of our new requirement that spectral characterization time be $<2$ months; this suggests that exoEarth spectral characterization for coronagraph-based missions doesn't ``turn on" until apertures $\sim$4 m.  Second, the dependence on IWA is diminished for the 8 m off-axis with VC and the 15 m on-axis with APLC; this implies that while a smaller IWA allows access to more stars, these missions do not have enough time to devote to those stars and are far into the time- or throughput-limited regimes.

\begin{figure}[H]
\centering
\includegraphics[width=\linewidth]{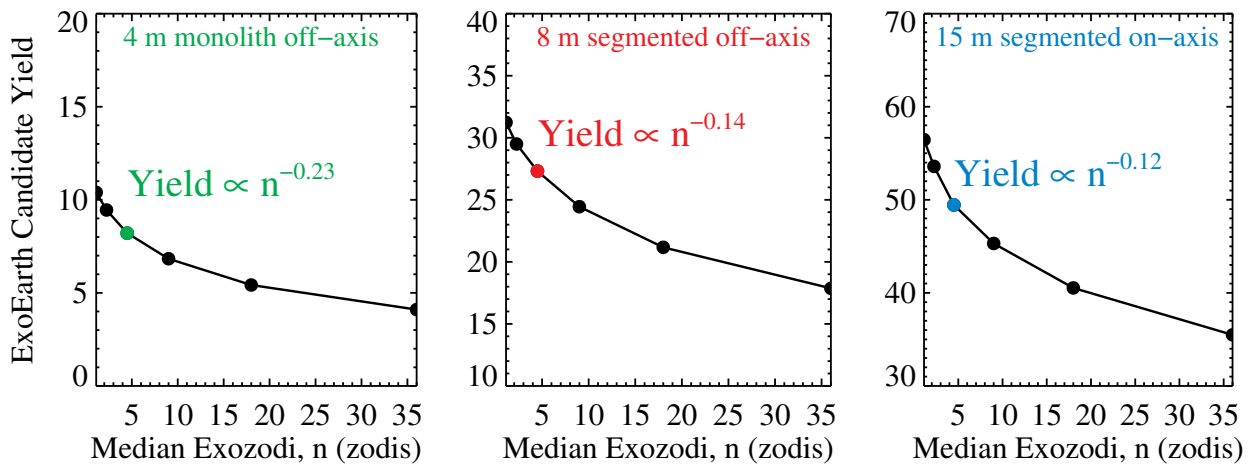}
\caption{The exoEarth candidate yield sensitivity to changes in median exozodi for the three coronagraph-based point designs shown in black in Figure \ref{yield_figure}. The largest aperture is least sensitive to exozodi.\label{exozodi_figure}}
\end{figure}

Figure \ref{exozodi_figure} shows the sensitivity of the three baseline coronagraph-based missions to median exozodi level.  Unlike the results of Ref.~\citenum{stark2014_2}, the sensitivity to exozodi decreases as aperture size increases. It is unclear whether this trend is caused by exposure time effects (e.g., due to adoption of a 2 month exposure time cutoff),  target selection effects (e.g., larger aperture's larger target list may provide more flexibility to selectively observe stars with low exozodi levels), or a combination of the two.

While a useful metric, the exoEarth yield shown in Figure \ref{yield_figure} does not tell the full story of the scientific return for each of the missions simulated.  There are additional differences to the instruments studied here to note. Because of the differences in optical performance and efficient operation of starshades and coronagraphs, they will produce fundamentally different data sets.  Starshade-based missions would produce data sets with many spectra over a wide wavelength range and large field of view, but with relatively few phase variations and orbit measurements.  Coronagraph-based missions, on the other hand, would produce the opposite---data sets with many detections, phase variations, measured orbits, and small spectral ``snippets," but a fraction of planets with spectra over a wide wavelength range.

Further, there are differences among the coronagraph-based missions for each of the three OTA scenarios studied here.  Specifically, the fields of view of the VC and APLC are very different. The APLC masks designed here are limited to OWA/IWA $\lesssim3$.  This generates a fairly narrow field of view that is optimized for the viewing of HZs.  The VC and DMVC would have a much larger field of view.  To obtain equally ``complete" views of a planetary system, the APLC would require 1--2 additional exposures depending on telescope diameters and science requirements. Additionally, the bandwidth of the VC is larger than the most aggressive APLC.  Larger bandwidths contribute to larger yields by reducing the time to detect planets in broadband, a factor that is already taken into account in Figure \ref{yield_figure}.  However, larger bandwidths also provide more complete spectra in a single characterization observation, a factor that does not play into the yield metric shown in Figure \ref{yield_figure}.  	

However, Figure \ref{yield_figure} does provide one clear conclusion.  To constrain the frequency of any feature in the spectra of exoEarth candidates, whether it be signs of habitability, the presence of biosignature gases, etc., Ref.~\citenum{stark2014_2} argued that a minimum compelling yield goal was $\sim$30 exoEarth candidates.  As shown in Figure \ref{yield_figure}, with the SAG13 occurrence rates and our adopted exoEarth candidate definition, there are two regions of design space where this may be possible: an off-axis segmented telescope with inscribed diameter $\gtrsim7$ m, and an on-axis segmented telescope with inscribed diameter $\gtrsim10$ m.

\section{Conclusions}
\label{conclusions}

Using improved yield models with high-fidelity simulations of realistic coronagraph and starshade designs, including their sensitivity to stellar diameter, we calculated the exoEarth candidate yield as a function of telescope diameter to assess several major telescope and instrument design trades.  We confirm previous results showing that in the absence of refueling, the yields of starshade-based missions are limited due to fuel use, while the yields of coronagraph-based missions continue to increase with telescope diameter. We find that for coronagraph-based missions, compared to monolithic off-axis telescopes there appears to be no yield penalty for segmentation if 1) the gap size between segments can be sufficiently small, such that the DMs can recreate a monolithic pupil, 2) we consider only the inscribed diameter of the primary mirror, and 3) the engineering tolerances---notably on wavefront control and stability---can be met in both scenarios.  We find that for coronagraph-based missions, there is a yield penalty for going to an on-axis segmented telescope design, but this can be overcome by going to even larger diameters.  Coronagraph design is an active area of research and we expect future advancements to improve their compatibility with on-axis telescopes.

\acknowledgments

This work was supported in part by the NASA Exoplanet Exploration Program's (ExEP) Segmented-aperture Coronagraph Design and Analysis (SCDA) study, the HabEx design study, the LUVOIR design study, and the WFIRST CGI Science Investigation Team contract \#NNG16P27C (PI: Margaret Turnbull). The results reported herein benefitted from collaborations and/or information exchange within NASA's Nexus for Exoplanet System Science (NExSS) research coordination network sponsored by NASA's Science Mission Directorate. This research has made use of the SIMBAD database, operated at CDS, Strasbourg, France. This work presents results from the European Space Agency (ESA) space mission Gaia. Gaia data are being processed by the Gaia Data Processing and Analysis Consortium (DPAC). Funding for the DPAC is provided by national institutions, in particular the institutions participating in the Gaia MultiLateral Agreement (MLA). J.~M. acknowledges support for this work was provided by NASA through the NASA Hubble Fellowship grant \#HST-HF2-51414 awarded by the Space Telescope Science Institute, which is operated by the Association of Universities for Research in Astronomy, Inc., for NASA, under contract NAS5-26555.

\bibliography{ms_v3.bbl}

\listoffigures
\listoftables

\end{spacing}
\end{document}